\documentclass[twocolumn,english,aps,pra,eqsecnum]{revtex4-1}
\usepackage[T1]{fontenc}
\usepackage[latin9]{inputenc}
\usepackage{color}
\usepackage{array}
\usepackage{float}
\usepackage{bm}
\usepackage{amsmath}
\usepackage{amssymb}
\usepackage{graphicx}
\usepackage{esint}

\makeatletter

\providecommand{\tabularnewline}{\\}
\newcommand{\lyxdot}{.}

\makeatother

\usepackage{babel}
\begin{document}

\title{Creation, storage and retrieval of an optomechanical cat state}

\author{R. Y. Teh, S. Kiesewetter}

\affiliation{Centre for Quantum and Optical Science, Swinburne University of Technology,
Melbourne 3122, Australia}

\author{P. D. Drummond and M. D. Reid}

\affiliation{Centre for Quantum and Optical Science, Swinburne University of Technology,
Melbourne, Australia}

\affiliation{Institute of Theoretical Atomic, Molecular and Optical Physics,Harvard
University, Cambridge, Massachusetts 02138, USA.}
\begin{abstract}
We analyze a method for the creation, storage and retrieval of optomechanical
Schr\"odinger cat states, in which there is a quantum superposition
of two distinct macroscopic states of a mechanical oscillator. In
the quantum memory protocol, an optical cat state is first prepared
in an optical cavity, then transferred to the mechanical mode, where
it is stored and later retrieved using control fields. We carry out
numerical simulations for the quantum memory protocol for optomechanical
cat states using the positive-P phase space representation. This has
a compact, positive representation for a cat state, thus allowing
a probabilistic simulation of this highly non-classical quantum system.
It is essential to use importance sampling to carry out the simulation
effectively. To verify the effectiveness of the cat-state quantum
memory, we consider several cat-state signatures and show how they
can be computed. We also investigate the effects of decoherence on
a cat state by solving the standard master equation for a simplified
model analytically, allowing us to compare with the numerical results.
Focusing on the negativity of the Wigner function as a signature of
the cat state, we evaluate analytically an upper bound on the time
taken for the negativity to vanish, for a given temperature of the
environment of the mechanical oscillator. We show consistency with
the numerical methods. These provide exact solutions, allowing a full
treatment of decoherence in an experiment that involves creating,
storing and retrieving mechanical cat states using temporally mode-matched
input and output pulses. Our analysis treats the internal optical
and mechanical modes of an optomechanical oscillator, and the complete
set of input and output field modes which become entangled with the
internal modes. The model includes decoherence due to thermal effects
in the mechanical reservoirs, as well as optical and mechanical losses.
\end{abstract}
\maketitle

\section{introduction}

Schr\"odinger's cat \citep{schrodinger1935gegenwartige} features
in the investigation of a fundamental issue in quantum mechanics \citep{RevModPhys.85.1083,RevModPhys.85.1103,Arndt:2014aa}
namely: Does quantum mechanics hold true in the macroscopic world?
This highly nonclassical state is also potentially useful, being proposed
as a resource in many quantum information applications including quantum
computation \citep{PhysRevA.59.2631,PhysRevA.61.042309,0953-4075-45-18-185505},
quantum teleportation \citep{PhysRevA.64.022313}, quantum metrology
\citep{PhysRevA.49.2151,PhysRevLett.107.083601} and cryptography
\citep{PhysRevA.89.012315}. As such, there has been much interest
in creating Schr\"odinger cat states of increasing size \citep{Ourjoumtsev83,PhysRevLett.97.083604,Wakui:07,Ourjoumtsev:2007aa,Deleglise:2008aa,PhysRevLett.101.233605,Ourjoumtsev:2009aa,PhysRevA.82.031802,Bimbard:2010aa,Yukawa:13,Dong:14,PhysRevLett.114.193602,PhysRevLett.115.023602,Ulanov:2016aa}.
Recent experiments use superconducting qubits to generate a cat state
that is a superposition of two distinguishable coherent states, with
the square distance in phase space between the two coherent states
up to 80 photons \citep{Wang1087} and more recently, 100 photons
\citep{Vlastakis607}. It remains a challenge however to prepare
a massive, \emph{mechanical} system in a cat state, which has the
potential for testing theories of quantum gravity. 

As well as being of fundamental importance, there are potential applications.
In a proposed quantum internet \citep{PhysRevLett.78.3221,kimble2008quantum,ritter2012elementary},
information is transmitted by light in a network of nodes connected
by optical fibers. At each node, the quantum information is received
and stored, to be later read out or sent to other nodes. A quantum
memory \citep{Duan:2001aa,PhysRevA.79.022310} is then essential as
the information of a quantum state needs to be stored on demand. An
optomechanical system is a good candidate for a quantum memory, where
the quantum state is stored in long-lived mechanical modes. In an
optomechanical system, the optical and mechanical modes have been
demonstrated to interact via radiation pressure in such a way that
state transfer between these modes is achievable \citep{palomaki2013coherent}.
In this work, we investigate the storage of a cat state in an optomechanical
system. We consider cat states that are a superposition of two distinguishable
coherent states.

There have been several earlier proposals to create cat states in
mechanical systems. This is a timely goal as quantum control in optomechanics
has dramatically improved, notably with the experimental observations
of ground state cooling \citep{schliesser2008resolved,teufel2011sideband,chan2011laser},
quantum state transfer \citep{palomaki2013coherent,Reed:2017aa} and
entanglement generation \citep{Palomaki710,Wang1087,Riedinger:2018aa,Ockeloen-Korppi:2018aa}
to name a few. In the case of optomechanical cat-state generation,
highly nonlinear interactions are typically required. Recently, there
are novel schemes to create \citep{PhysRevLett.117.143601,PhysRevX.8.021052}
and enlarge the size of optomechanical cat states \citep{Sychev:2017aa,1367-2630-20-5-053042,2058-9565-4-1-014003}. 

Here, we consider an alternative method that involves quantum state
transfer from an external optical cavity to the mechanical system,
which is essentially utilized as a quantum memory. The type of quantum
memory utilized here is an on-demand synchronous dynamical memory,
in which the mode-matched input and output of the memory is facilitated
by the use of shaped gain and detuning, as treated in previous mode-matched
intracavity quantum memory proposals \citep{PhysRevA.79.022310,he2009digital}.
This general strategy has been previously analyzed for generation
of entangled mechanical states \citep{PhysRevLett.119.023601}. There
are related proposals suggested for systems without cavities \citep{PhysRevA.68.013808,PhysRevLett.105.070403}
and some recent strategies in optomechanics of a similar nature, but
with different protocols \citep{1367-2630-13-1-013017,PhysRevLett.108.153603}. 

In our proposal, an optical cat-state is prepared externally, transferred
and stored as a mechanical cat-state. It is later retrieved on demand
using control fields. An advantage of this method is that optical
or microwave cat-states have been generated with high fidelity \citep{Ourjoumtsev:2007aa,Vlastakis607}.
The storage time is completely controllable, allowing an analysis
of decoherence effects. Finally, the verification measurements can
be made externally, using well-developed optical homodyne techniques.
This is essential, as there are no current techniques that would allow
an\emph{ in-situ }quantum state tomography of a mechanical oscillator.

For an efficient quantum memory, the coupling between the input state
and the physical system has to be optimized. The system also has to
store a quantum state in the desired mode. These goals are achieved
with mode matching by choosing an optimal mode function. In Section
$II$, we provide a description of a protocol using mode matching
for transferring the cat-state between the optical and mechanical
modes. The protocol involves the storage and retrieval of the state,
as in a quantum memory. A theoretical model for this protocol was
developed earlier \citep{PhysRevA.96.013854}. That work however only
examined the transfer of a coherent state. The model included dissipation
as well as thermal noise. 

A cat state is sensitive to fluctuations and noise due to the interaction
with its environment. Hence, measurable signatures are needed to verify
the presence of a cat state. In Section $III$, we summarize several
quantities that might be used to signify a cat state. These quantities
can then be used to give a measure of the effectiveness of the cat-state
storage and retrieval. The signatures we examine are the fringe patterns
in quadrature probability distributions \citep{PhysRevLett.117.143601,PhysRevX.8.021052},
the Wigner function \citep{PhysRevLett.89.200402,Deleglise:2008aa,Vlastakis607,Wang1087}
and its negativity \citep{PhysRevLett.117.143601,2058-9565-4-1-014003},
the off-diagonal terms of the density operator \citep{Deleglise:2008aa},
and a variance signature \citep{PhysRevLett.97.170405,PhysRevA.76.030101,PhysRevA.77.062108,1367-2630-14-9-093039,PhysRevA.89.012116,Oudot:15,FROWIS20152,PhysRevA.94.062125,PhysRevLett.116.090801}. 

We also give an analytical treatment of the decoherence of an idealized
cat-state, with detailed calculations of this simplified model in
the Appendix, taking into account the thermal occupation number $\bar{n}_{th}$
of the mechanical oscillator reservoir, by solving the standard master
equation. The solution tells us how off-diagonal terms decay in time
as a function of the cat size, and also provides a way to calculate,
for a given value of $\bar{n}_{th}$, an upper bound on the time for
a Wigner function to lose its negativity, which is a typical indicator
of non-classicality. 

As discussed by Paavola et al. \citep{PhysRevA.84.012121}, a ``sudden
death'' effect is observed in the presence of thermal noise, which
fully converts the cat state to a mixture if the cat state is coupled
to the thermal reservoir for long enough time. We report however that
the first two signatures undergo a \emph{premature} ``sudden death''
effect for sufficient coupling time in the absence of thermal noise,
failing to indicate non-classicality despite that the cat state has
not fully decohered to a mixture of coherent states. 

In Section $IV$ the positive-P phase space method used to solve the
full quantum memory model is explained. Despite the highly nonclassical
states involved, this can be readily achieved using an exact probabilistic
mapping of the full quantum state into a phase-space representation.
This involves numerical simulation via the positive-P representation
\citep{0305-4470-13-7-018}, where the dimensionality of the complex
phase space is doubled. In this approach the entire input-output process,
including all participating optical and mechanical modes as well as
dissipation and noise are included, in an exact quantum dynamical
simulation. The cat state can be easily treated using this method
if an appropriate importance sampling technique is used. This section
deals with the methodology for the sampling of the cat-state and its
dynamics. 

The results of our simulations including predictions for the cat-state
signatures and a discussion of feasibility is given in Section $V$.
Here we use typical parameter values from recent electromechanical
experiments. We analyze in detail the effects of the different types
of decoherence present in the full model. This treats the complete
protocol starting from an externally generated cat state, storing
it in a mechanical mode, then retrieving and analyzing it externally.
As expected, the greater the level of loss and decoherence present,
the more quickly the cat signatures are extinguished. We find that
cat states with up to $9$ mechanical phonons can be stored for short
periods with realistic parameter values corresponding to current experiments.
This corresponds to a distance metric of $S=\left|\alpha_{1}-\alpha_{2}\right|^{2}=36$.
Further improvements in temperature and loss rates will be needed
to reach $S=100,$ which is the largest cat state generated to date
using superconducting waveguide techniques \citep{Vlastakis607}.
Results are summarized in Section VI.

\section{cat-state transfer }

\subsection{Cat-state generation}

In electro-optical experiments, cat states have been generated at
microwave frequency inside a cavity \citep{Vlastakis607}. We consider
the cat state as a quantum superposition of two coherent states $|\alpha_{0}\rangle$
and $|-\alpha_{0}\rangle$, in a mode with a corresponding operator
$a_{0}\left(t\right)$. This original idealized cat-state has the
form
\begin{align}
|\psi_{cat}\rangle & =\frac{1}{\sqrt{\mathcal{N}}}\left(|\alpha_{0}\rangle+|-\alpha_{0}\rangle\right)\,,\label{eq:cat_state}
\end{align}
where the normalizing factor is:
\begin{equation}
\mathcal{N}\equiv2\left(1+\text{exp}\left(-2\left|\alpha_{0}\right|^{2}\right)\right)\,.
\end{equation}
We note that this state will not be completely ideal due to losses
and thermal noise, but we assume here that we start with an idealized
cat state, in order to analyze the storage and retrieval process.

Having been generated, the state is assumed to be rapidly out-coupled
to a waveguide, on time-scales that are short compared to the originating
cavity internal losses and nonlinearities. Following a generic model
given in previous work \citep{PhysRevA.79.022310,he2009digital,PhysRevLett.119.023601},
we assume that the output coupler is time-dependent. Using input-output
theory, one therefore obtains:
\begin{align}
\frac{d}{dt}a_{0}\left(t\right) & =-\kappa\left(t\right)a_{0}\left(t\right)+\sqrt{2\kappa\left(t\right)}\hat{\phi}_{0}^{in}\nonumber \\
\hat{\phi}_{0}^{out} & =\sqrt{2\kappa\left(t\right)}a_{0}-\hat{\phi}_{0}^{in}\,.\label{eq:in-out}
\end{align}

We assume that the state is prepared at time $t=t_{0}=-t_{W}$, then
out-coupled at times $t>-t_{W}$, by turning on the output coupler
so that $\kappa\left(t\right)>0$, up until the end of the output
process at $t=0$. The resulting solution for $a_{0}\left(t\right)$
is therefore:
\begin{equation}
a_{0}(t)=e^{-K(t)}\left[a_{0}\left(t_{0}\right)+\int_{t_{0}}^{t}e^{K(\tau)}\sqrt{2\kappa\left(\tau\right)}\hat{\phi}_{0}^{in}\left(\tau\right)d\tau\right]\,,
\end{equation}
where, 
\begin{equation}
K(t)=\int_{t_{0}}^{t}\kappa\left(\tau\right)d\tau
\end{equation}

We choose $K(t)$ and hence $\kappa\left(\tau\right)$ so that it
corresponds to a precise, temporally mode-matched function $u_{0}\left(t\right)$,
where we defube $u_{0}$ such that $\hat{\phi}_{0}^{out}\left(t\right)=u_{0}\left(t\right)a_{0}\left(t\right)+\text{noise}$,
which implies 
\begin{equation}
u_{0}\left(t\right)=\sqrt{2\kappa\left(t\right)}\exp\left(-\int_{t_{0}}^{t}\kappa\left(\tau\right)d\tau\right)\label{eq:temp_mode}
\end{equation}
This requires that $\kappa\left(\tau\right)$ obeys the following
equation:
\begin{equation}
\frac{d}{dt}\kappa(t)=2\kappa(t)\frac{d}{dt}\ln u_{0}\left(t\right)+2\kappa^{2}(t).
\end{equation}
In general, this is a nonlinear differential equation that requires
a numerical solution for any given mode-matching requirement. There
are special cases that are analytically soluble, however. One simple
case is for a rising exponential mode-function. This is a case that
corresponds to the required mode-matched input in the present situation,
to a good approximation as described later, with:
\begin{equation}
u_{0}(t)=\sqrt{2\bar{\gamma}}\exp(\bar{\gamma}t)\Theta\left(-t\right)\,.
\end{equation}
Here, $\bar{\gamma}$ is a parameter that sets the time scale of the
state transfer protocol as described later. The output coupler solution
is then, in the limit of $-\bar{\gamma}t_{0}\gg1$, and for $t<0$,
\begin{equation}
\kappa(t)=\bar{\gamma}\left(e^{-2\bar{\gamma}t}-1\right)^{-1}.
\end{equation}

This solution is rather simple mathematically. However, it is not
the simplest to implement. A one-sided pulse-shape leads to a singular
coupling in this approximation, and this appears to be a generic issue
related to the sharp temporal cut-off used here in order to have well-defined
cat storage times. Yet smooth, two-sided solutions exist as well.
These are described in earlier work \citep{he2009digital,PhysRevLett.119.023601}.
The details of this type of mode implementation, and how these change
our results, will be given elsewhere. 

\subsection{Cat-state download protocol \label{subsec:Optomechanical-state-transfer}}

\begin{figure}[H]
\begin{centering}
\includegraphics[width=0.7\columnwidth]{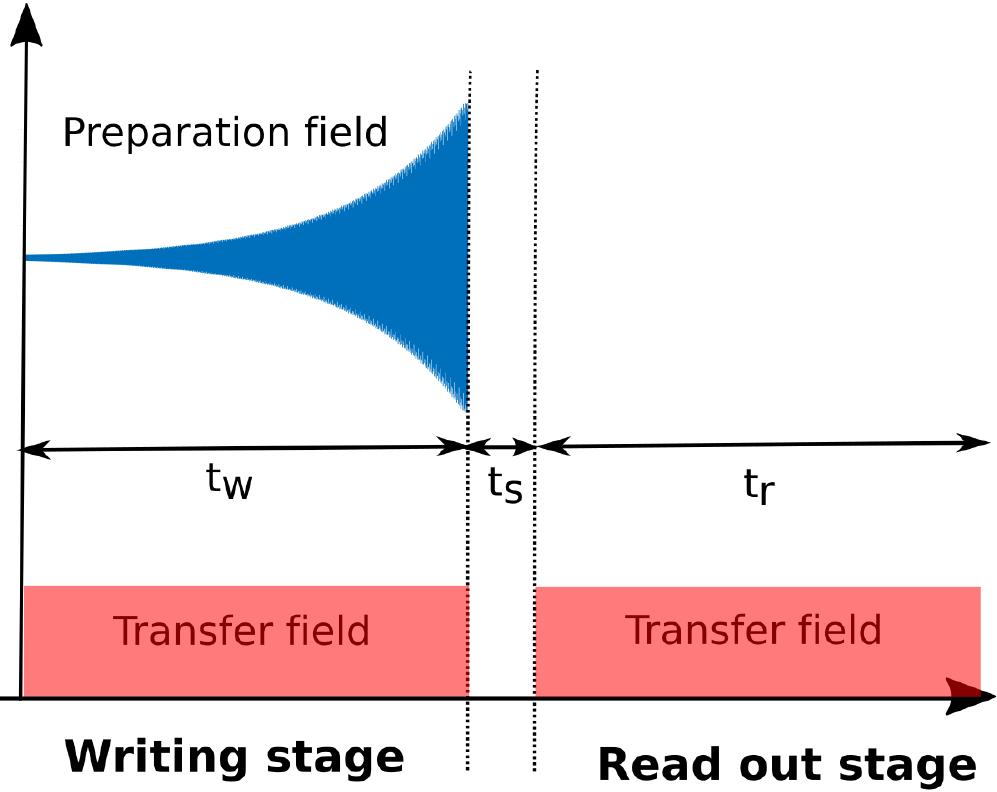}
\par\end{centering}
\caption{The state transfer protocol. During the writing stage, both the preparation
field containing the cat state, and the transfer field that couples
the preparation field to the cavity are turned on for a duration of
$t_{w}$. Both fields are turned off during the storage stage for
$t_{s}$. The stored state is read out by applying a second transfer
field for $t_{r}=t_{w}$. \label{fig:opto_transfer_protocol}}
\end{figure}

The approach to optomechanical state transfer used here is based on
previous work \citep{PhysRevA.96.013854}, which we indicate schematically
in Fig. \ref{fig:opto_transfer_protocol}.

The dynamical protocol for writing the input cat state into the mechanical
mode and later retrieving it, requires two pulses at each stage: the
preparation and transfer fields. The preparation field is resonant
to the optical cavity resonance frequency $\omega_{o}$, and it prepares
the optical cat-state in the cavity. The transfer field, with a frequency
$\omega_{d}$ such that the detuning between the cavity and transfer
field is the mechanical mode frequency $\Delta=\omega_{o}-\omega_{d}=\omega_{m}$,
facilitates the transfer of the cat-state between the optical and
mechanical modes. In total, the state transfer protocol consists of
three stages, as shown in Fig. \ref{fig:opto_transfer_protocol}.
The optical quantum state is first generated externally at time $t=t_{0}=-t_{W}$,
and transferred to the mechanical state of motion. We note that this
process is carried out using a temporal mode-matching protocol to
allow efficient transfer, leaving the external source cavity in a
vacuum state.

After the successful transfer, both fields are turned off for a time
$t_{s}$. This allows the system to store the mechanical cat for a
prescribed time. This interval needs to be made as long as possible,
in order to test decoherence theories. When the quantum state is to
be read out, the transfer field is turned on. The stored quantum state
is transferred back to an intra-cavity optical mode. Finally, it is
coupled out of the cavity with temporal mode-matching to a waveguide
traveling mode of duration $t_{r}$, for external detection. The protocol
is completed at the final time, $t=t_{f}=t_{s}+t_{r}$.

This quantum memory protocol \citep{PhysRevA.79.022310,he2009digital}
has been experimentally implemented \citep{palomaki2013coherent}
in a superconducting electromechanical device. It is a dynamical scheme
which transfers the prepared state from an external source to the
mechanical mode. Temporal mode-matching is used both for input and
output. This ensures efficient transfer to and from the external multi-mode
waveguide modes. The mechanical state can be coupled out after a well-defined
storage time. This procedure allows for studies of time-dependent
decoherence. 

\subsection{Quantum optomechanical Hamiltonian}

\begin{figure}[H]
\begin{centering}
\includegraphics[width=0.9\columnwidth]{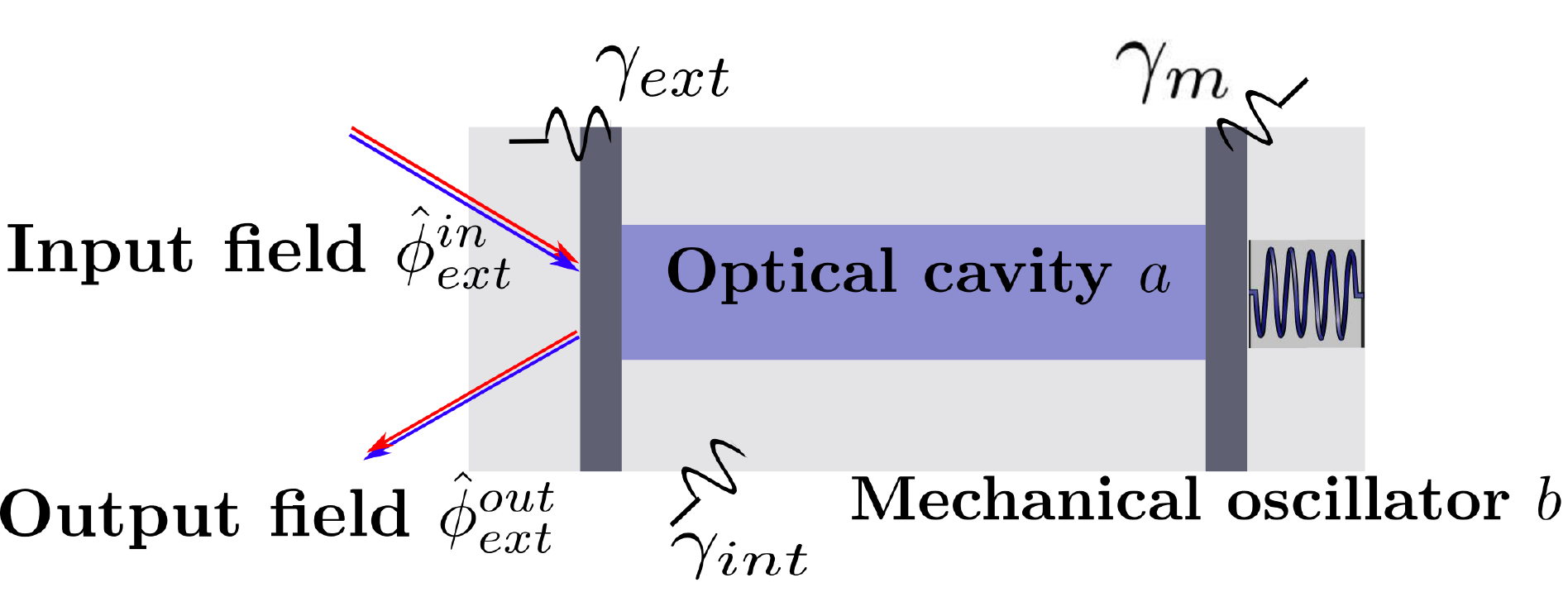}
\par\end{centering}
\caption{Schematic diagram of the optomechanical system. \label{fig:opto_transfer_schematic}}
\end{figure}

A typical optomechanical system consists of an optical cavity and
a mechanical oscillator that interact via radiation pressure as shown
in Fig. \ref{fig:opto_transfer_schematic}. The optics and mechanics
are characterized by their resonance frequencies and decay rates.
In the single mode model, the optical cavity and mechanical oscillator
have resonance frequencies $\omega_{o}$ and $\omega_{m}$ respectively;
other frequencies are not involved and can be neglected. 

The decay rate of the mechanical oscillator is $\gamma_{m}$ while
we identify two separate sources of dissipation in the optical cavity:
the \textit{internal} and \textit{external} decay rates, $\gamma_{int}$
and $\gamma_{ext}$. The total optical cavity decay rate is $\gamma_{o}=\gamma_{int}+\gamma_{ext}$.
The external cavity decay rate $\gamma_{ext}$ determines the coupling
strength of an input and output field to the cavity, which allows
the detection of the cavity optical field. All other sources of dissipation
are included in the internal decay rate $\gamma_{int}$. 

The dynamics of an optomechanical system is given by the following
standard Hamiltonian \citep{pace1993quantum,PhysRevLett.88.120401},
in the rotating frame of an external laser field,
\begin{eqnarray}
H & = & \hbar\Delta a^{\dagger}a+\hbar\omega_{m}b^{\dagger}b+\hbar g_{0}a^{\dagger}a\left(b+b^{\dagger}\right)\nonumber \\
 &  & +\hbar\epsilon\left(t\right)\left(a^{\dagger}+a\right)\,,\label{eq:typical Hamiltonian}
\end{eqnarray}
where $\Delta=\omega_{o}-\omega_{d}$ is the detuning between the
cavity resonance frequency $\omega_{o}$ and the external laser carrier
frequency $\omega_{d}$. The third term in Eq. (\ref{eq:typical Hamiltonian})
corresponds to the nonlinear radiation pressure interaction between
the optical and mechanical modes, with a coupling strength determined
by $g_{0}$. The mode operators $a,b$ correspond to the intracavity
optical and mechanical modes respectively. 

The last term includes all external fields $\epsilon\left(t\right)$
that are sent into the optical cavity, which includes the external
cat state that is imprinted into the system and the transfer field.
This is described in greater detail later. In this work, we describe
the radiation pressure interaction term using a simplified, linearized
optomechanical Hamiltonian in the interaction picture, where:
\begin{align}
H_{int} & =\hbar g(t)\left(ab^{\dagger}+a^{\dagger}b\right)\,.\label{eq:linearized_Hamiltonian}
\end{align}
Here $g(t)=\sqrt{N}g_{0}\left(\Theta\left(-t\right)+\Theta\left(t-t_{s}\right)\right)$
is the effective coupling strength and $N$ is the mean photon number
in the cavity due to the off-resonant transfer field used for switching
\citep{0295-5075-54-5-559,PhysRevLett.88.120401}, $\Theta$ is the
Heaviside step function, and $t_{s}$ is the storage time. A rotating
wave approximation is used in obtaining the interaction Hamiltonian
Eq. (\ref{eq:linearized_Hamiltonian}). The linearization approximation
is valid when an intense off-resonant driving field is applied to
enhance the optomechanical coupling strength, and when the cavity
decay rate is much smaller than the mechanical frequency (the resolved
sideband regime) \citep{PhysRevA.68.013808}. We have investigated
the validity of this approximation elsewhere, by carrying out full
nonlinear quantum simulations \citep{PhysRevA.90.043805,PhysRevA.96.013854}.

We treat the optomechanical system as an open quantum system, where
the fluctuations of the system due to interactions with its environment
are taken into account. This is necessary: a quantum cat-state is
fragile and sensitive to perturbations. A standard formalism for treating
such an open system is provided by the master equation \citep{carmichael2013statistical}:
\begin{eqnarray}
\frac{d}{dt}\hat{\rho} & = & -\frac{i}{\hbar}\left[H,\hat{\rho}\right]+\sum_{j}\gamma_{j}\left[\bar{n}_{j}\left(2a_{j}^{\dagger}\hat{\rho}a_{j}-a_{j}\hat{\rho}a_{j}^{\dagger}-\hat{\rho}a_{j}a_{j}^{\dagger}\right)\right.\nonumber \\
 &  & +\left.\left(\bar{n}_{j}+1\right)\left(2a_{j}\hat{\rho}a_{j}^{\dagger}-a_{j}^{\dagger}a_{j}\hat{\rho}-\hat{\rho}a_{j}^{\dagger}a_{j}\right)\right]\,.\label{eq:master equation}
\end{eqnarray}
Here, $\hat{\rho}$ is the density operator of the optomechanical
system, the index $j=1,\,2\sim o,m$ refer to the cavity and mechanical
modes respectively, and $\bar{n}_{j}$ is the average thermal occupation
number from interactions with their corresponding reservoirs. 

In our work we extend this approach to include the relevant input
and output modes used to create and retrieve the cat state. Owing
to its complexity, it is more convenient to solve this large dynamical
quantum system using an efficient positive-P phase-space representation.
This maps the relevant density matrix into a positive probability
distribution, and its dynamics into a numerically tractable set of
stochastic equations. We note that while one can integrate the full
set of nonlinear equations generated by the full Hamiltonian $H$,
as we have done elsewhere, in this paper we take an idealized case
where only the linearized equations obtained from $H_{int}$ are treated.

\subsection{Input-output relations}

The state transfer protocol relies on an optimal mode-matching \citep{PhysRevA.79.022310,PhysRevA.96.013854}
for efficient coupling and detection of both the input and output
fields, to and from the optical cavity, respectively. These fields
have to be integrated with their corresponding temporal modes $u_{in}\left(t\right)$
and $u_{out}\left(t\right)$, which are obtained by solving the time
evolution equations of the optical $a$ and mechanical $b$ modes. 

There are four relevant bosonic mode operators in the model, as well
as an infinite set of \emph{`modes of the universe}' in the input
and output channels, giving a total Hilbert space of $\mathcal{H}$.
Apart from selected mode-matched input and output modes, these are
optimally maintained in a vacuum state to get the best fidelity, although
our method can treat other possibilities, and thermal phonon excitation
will be included.

The operator time evolution equations are quantum Langevin equations
obtained using the linearized optomechanical Hamiltonian in Eq. (\ref{eq:linearized_Hamiltonian}),
which are:
\begin{align}
\frac{d}{dt}a\left(t\right) & =-\gamma_{0}a-ig\left(t\right)b+\sqrt{2\gamma_{ext}}\hat{\phi}_{ext}^{in}+\sqrt{2\gamma_{int}}\hat{\phi}_{int}^{in}\nonumber \\
\frac{d}{dt}b\left(t\right) & =-\gamma_{m}b-ig\left(t\right)a+\sqrt{2\gamma_{m}}\hat{\phi}_{m}^{in}\,.\label{eq:quantum_langevin}
\end{align}
The total cavity decay rate is given by $\gamma_{0}=\gamma_{ext}+\gamma_{int}$,
where $\gamma_{ext}$ corresponds to output coupling losses through
the external mirrors and $\gamma_{int}$ corresponds to the remaining
internal losses in the cavity. The internal fields $\hat{\phi}_{int}^{in},\,\hat{\phi}_{m}^{in}$
are the quantum Langevin noise operators due to interaction of the
optomechanical system with its internal lossy environment for the
photons and mechanical phonons respectively. 

The following external input-output relation must also be satisfied:
\begin{equation}
\hat{\phi}_{ext}^{out}\left(t\right)=\sqrt{2\gamma_{ext}}a\left(t\right)-\hat{\phi}_{ext}^{in}\left(t\right)\,,
\end{equation}
where the external input and output fields are traveling waves. These
have a mode expansion for the field at the interface mirror given
by:
\begin{align}
\hat{\phi}_{ext}^{in}\left(t\right) & =\sum_{n\ge0}a_{n}^{in}u_{n}^{in}\left(t\right)\nonumber \\
\hat{\phi}_{ext}^{out}\left(t\right) & =\sum_{n\ge0}a_{n}^{out}u_{n}^{out}\left(t\right)\,.\label{eq:Externalfieldmodes}
\end{align}

Here $\hat{\phi}_{ext}^{in}$ is the external input into the cavity,
where $a_{0}^{in}$ is a mode operator for the source of the cat-state,
and $a_{n>0}$ is the set of external vacuum mode operators with orthogonal
temporal modes given by $u_{n}^{in}$. We wish to store the input
state of $a_{0}^{in}$ internally in the optomechanical device, where
$u_{0}^{in}\left(t\right)$ is the temporal mode of this preferred
input state. This is typically created in a second, external photonic
cavity or waveguide \citep{PhysRevLett.119.023601}, and transferred
on demand to the quantum memory, with an engineered temporal mode
shape. There are many proposals for creating such cat states in the
external cavity \citep{wolinsky1988quantum,reid1993macroscopic,krippner1994transient},
and this choice is left open here. In this work, we assume perfect
optomechanical input coupling from the source cavity, so $\hat{\phi}_{0}^{out}$
in Eq. (\ref{eq:in-out}) is equal to $\hat{\phi}_{ext}^{in}$ in
Eq. (\ref{eq:Externalfieldmodes}), and $u_{0}$ in Eq. (\ref{eq:temp_mode})
is equal to $u_{0}^{in}$ in Eq. (\ref{eq:in-out}). There is also
an output mode $\hat{\phi}_{ext}^{out}\left(t\right)$, defined by
the the input-output relation \citep{PhysRevA.31.3761} given above.

These equations are based on the input-output formalism developed
by Gardiner and Collett \citep{PhysRevA.31.3761}. Similar treatments
of the quantum nature of the optomechanical coupling for the study
of entanglement have been given by Hofer et al. \citep{PhysRevA.84.052327},
He and Reid \citep{PhysRevA.88.052121}, and Kiesewetter et al. \citep{PhysRevA.90.043805,PhysRevLett.119.023601}. 

\subsection{Optimized mode function}

Details of the calculations and derivations of these temporal modes
can be found in the work of Teh et al. \citep{PhysRevA.96.013854}.
Here, we note that the solutions of the quantum Langevin equations
in Eq. (\ref{eq:quantum_langevin}) are obtained. From these solutions,
the optimal temporal mode function $u_{0}^{in}\left(t\right)$ that
gives the best mode-matching - in terms of transfer efficiency - is
found to be
\begin{align}
u_{0}^{in}\left(t\right) & =-2i\frac{\sqrt{\left(\gamma_{+}+m\right)\left(\gamma_{+}-m\right)\gamma_{+}}}{m}\text{sinh}\left(mt\right)e^{\gamma_{+}t}\Theta(-t)\,,\nonumber \\
\label{eq:input_mode_function}
\end{align}
where $\gamma_{+}=\left(\gamma_{o}+\gamma_{m}\right)/2$, $\gamma_{-}=\left(\gamma_{o}-\gamma_{m}\right)/2$,
$m=\sqrt{\gamma_{-}^{2}-g^{2}}$, $g=\sqrt{N}g_{0}$ is the effective
optomechanical coupling strength, and $\Theta$ is the Heaviside step
function. Here we assume that $N\left(t\right)=N\Theta\left(-t\right)$.
The corresponding output temporal mode function $u_{0}^{out}\left(t\right)$
is related to the input temporal mode function $u_{0}^{in}\left(t\right)$
via $u_{0}^{out}\left(t\right)=u_{0}^{in*}\left(t_{s}-t\right)$,
with $N\left(t\right)=N\Theta\left(t-t_{s}\right)$. 

In particular, the stored mode operator is 
\begin{align}
b\left(0\right) & =\frac{\sqrt{2\gamma_{ext}}ga_{0}}{2\sqrt{\left(\gamma_{+}+m\right)\left(\gamma_{+}-m\right)\gamma_{+}}}+\text{noise}\,.\label{eq:stored_mode}
\end{align}
From orthonormality of the relevant mode functions, the mode input
$a_{0}^{in}$ and output $a_{0}^{out}$ containing the fields to be
stored and retrieved, respectively, in the optomechanical system are
given by:
\begin{eqnarray}
a_{0}^{in} & = & \intop_{-\infty}^{0}u_{0}^{in*}\left(t\right)\hat{\phi}_{ext}^{in}\left(t\right)\,dt\nonumber \\
a_{0}^{out} & = & \intop_{t_{s}}^{\infty}u_{0}^{out*}\left(t\right)\hat{\phi}_{ext}^{out}\left(t\right)\,dt\,,\label{eq:truncated_wigner_integratedinout}
\end{eqnarray}
where $\hat{\phi}_{ext}^{in}\left(t\right)$, $\hat{\phi}_{ext}^{out}\left(t\right)$
are the cavity input and output fields. We note that, to a good approximation,
if $\gamma_{m}\ll g\ll\gamma_{o}$, if $\bar{\gamma}=\gamma_{+}-m$,
then:
\begin{equation}
u_{0}^{in}\left(t\right)\approx i\sqrt{2\bar{\gamma}}e^{\bar{\gamma}t}\Theta(-t)\,.
\end{equation}
 Apart from the phase-factor, which is readily adjustable, this is
the approximate exponential form analyzed in treating the download
phase from the original cavity. However, we use the full expression
in the numerical simulations.

\section{Cat-state Signatures \label{sec:cat_state_signatures}}

As a preliminary exercise, we first consider the signatures of a cat
state generated in a single stationary bosonic mode, which is a simplified
model of the mechanical mode. The corresponding density operator for
the cat state $\hat{\rho}_{cat}$ is then 
\begin{align}
\hat{\rho}_{cat} & =\frac{1}{\mathcal{N}}\left(|\alpha_{0}\rangle\langle\alpha_{0}|+|-\alpha_{0}\rangle\langle-\alpha_{0}|\right.\nonumber \\
 & \left.+|\alpha_{0}\rangle\langle-\alpha_{0}|+|-\alpha_{0}\rangle\langle\alpha_{0}|\right)\,.\label{eq:cat_state_density_op}
\end{align}

It is necessary to verify that the cat state is created and successfully
stored in a mechanical mode. This is done by verifying the strength
of the cat signature in the retrieved output mode after a storage
time $t_{s}$. In this paper, three possible cat state signatures
are investigated. One of the earliest signatures proposed in the literature
is the presence of interference fringes in the quadrature probability
density distribution \citep{PhysRevLett.57.13}. A second signature
is the negativity of the Wigner function, which can be quantified
by the negative volume of that Wigner function \citep{1464-4266-6-10-003}.
As a third signature, we reconstruct the density operator and infer
the presence of the optomechanical cat state from the off-diagonal
terms \citep{Deleglise:2008aa}. Finally, we discuss a novel variance
inequality cat signature, which when violated, implies that the physical
state is not in a mixture of two distinguishable coherent states. 

\subsection{Interference fringes in quadrature probabilities}

Using homodyne detection, the quadrature phase amplitudes can be measured,
after the state is transferred to an output photonic mode.  The interference
fringes in the quadrature probability distribution have been quantified
as a cat-measure \citep{PhysRevLett.117.143601,PhysRevX.8.021052}.
Generally, the rotated orthogonal quadratures $\hat{X}_{\theta}$
and $\hat{X}_{\theta+\frac{\pi}{2}}$ are defined in terms of creation
and annihilation operators as
\begin{align}
\hat{X}_{\theta} & =\frac{1}{\sqrt{2}}\left(e^{-i\theta}a+e^{i\theta}a^{\dagger}\right)\nonumber \\
\hat{X}_{\theta+\frac{\pi}{2}} & \equiv P_{\theta}=\frac{1}{i\sqrt{2}}\left(e^{-i\theta}a-e^{i\theta}a^{\dagger}\right)\,.\label{eq:rotated_quadrature}
\end{align}
 The inner product of a coherent state $|\alpha\rangle$ and a rotated
quadrature basis state $|x_{\theta}\rangle$, which is the eigenstate
of the quadrature operator $\hat{X}_{\theta}$ and satisfies $\hat{X}_{\theta}|x_{\theta}\rangle=x_{\theta}|x_{\theta}\rangle$,
can be shown to be given by \citep{PhysRevLett.57.13}
\begin{align}
\langle x_{\theta}|\alpha\rangle & =\frac{1}{\pi^{\frac{1}{4}}}exp[-\frac{x_{\theta}^{2}}{2}+\sqrt{2}e^{-i\theta}x_{\theta}\alpha-\frac{e^{-2i\theta}\alpha^{2}}{2}-\frac{|\alpha|^{2}}{2}]\,,\nonumber \\
\label{eq:general_rep_coherent-1}
\end{align}
with $\alpha=\left|\alpha\right|e^{i\phi}$ defined as the complex
amplitude of the coherent state $|\alpha\rangle$. In particular,
we will consider the case $\theta=0$, $x_{\theta=0}=x$ and $p_{\theta=0}=x_{\frac{\pi}{2}}=p$.
Without losing generality, we also consider a real coherent state
amplitude, setting $\phi=0$. 
\begin{figure}[H]
\begin{centering}
\includegraphics[width=0.8\columnwidth]{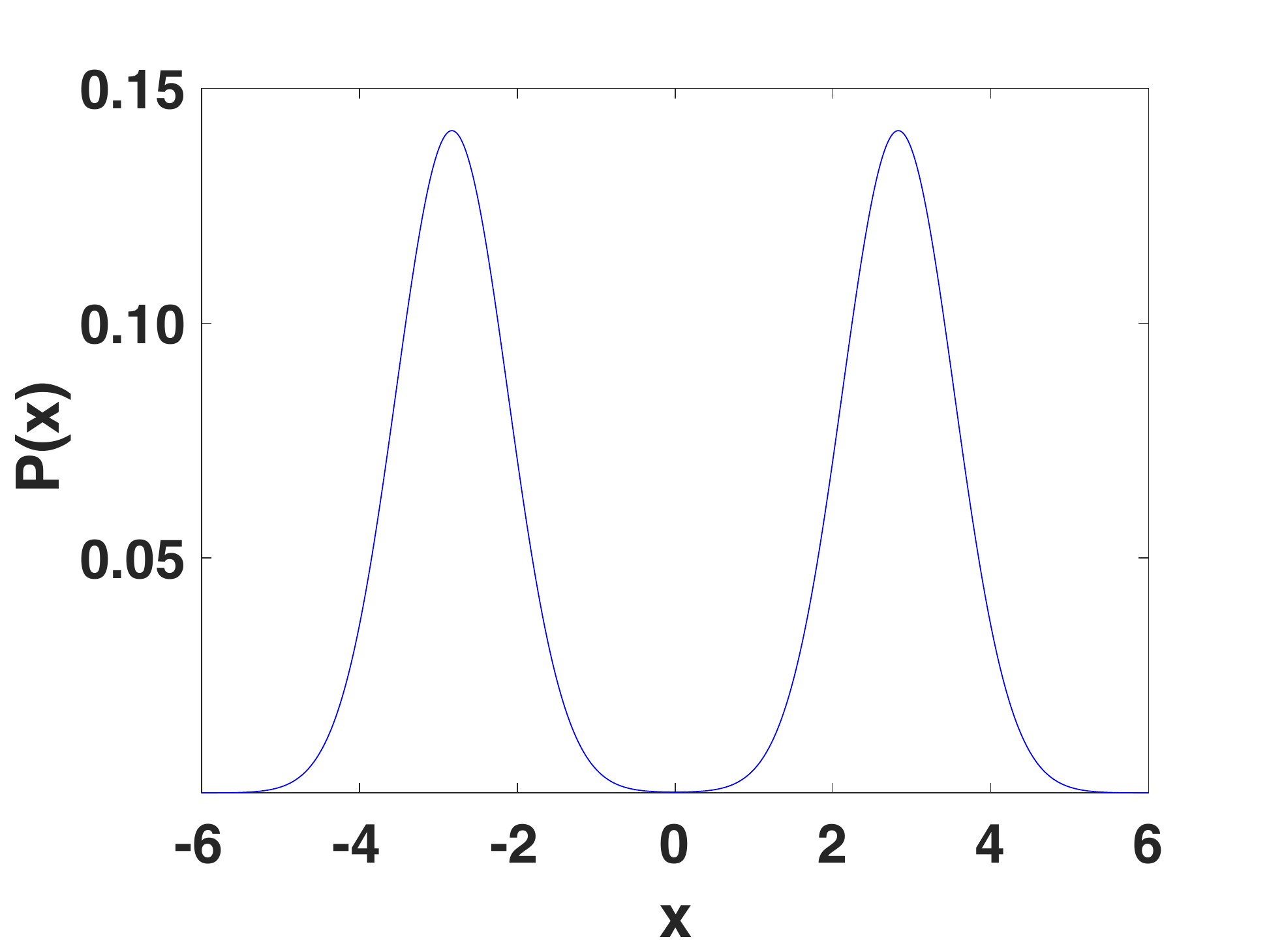}
\par\end{centering}
\caption{\label{fig:Probability-density-distribution-x}Probability density
distribution $P(x)$ for the $x$ quadrature of the cat state Eq.
(\ref{eq:cat_state}) with $\alpha_{0}=2$ as given in Eq. (\ref{eq:P(x)-distribution}).
\textcolor{red}{}}
\end{figure}

The corresponding probability distribution, following Eq. (\ref{eq:general_rep_coherent-1}),
is then

\begin{align}
P\left(x\right) & =\langle x|\hat{\rho}_{cat}|x\rangle\label{eq:P(x)-distribution}\\
 & =\frac{1}{\sqrt{\pi}\mathcal{N}}\left\{ exp\left[-\left(x-\sqrt{2}\alpha_{0}\right)^{2}\right]\right.\nonumber \\
 & \left.+exp\left[-\left(x+\sqrt{2}\alpha_{0}\right)^{2}\right]+2exp\left[-x^{2}-2\alpha_{0}^{2}\right]\right\} \nonumber 
\end{align}
for the $x$ quadrature. This distribution, $P\left(x\right)$, consists
of two exponential terms that correspond to two Gaussian hills around
the values $x=\pm\sqrt{2}\alpha_{0}$, and also a rapidly decaying
exponential term, as shown in Fig. \ref{fig:Probability-density-distribution-x}.

\begin{figure}[H]
\begin{centering}
\includegraphics[width=0.8\columnwidth]{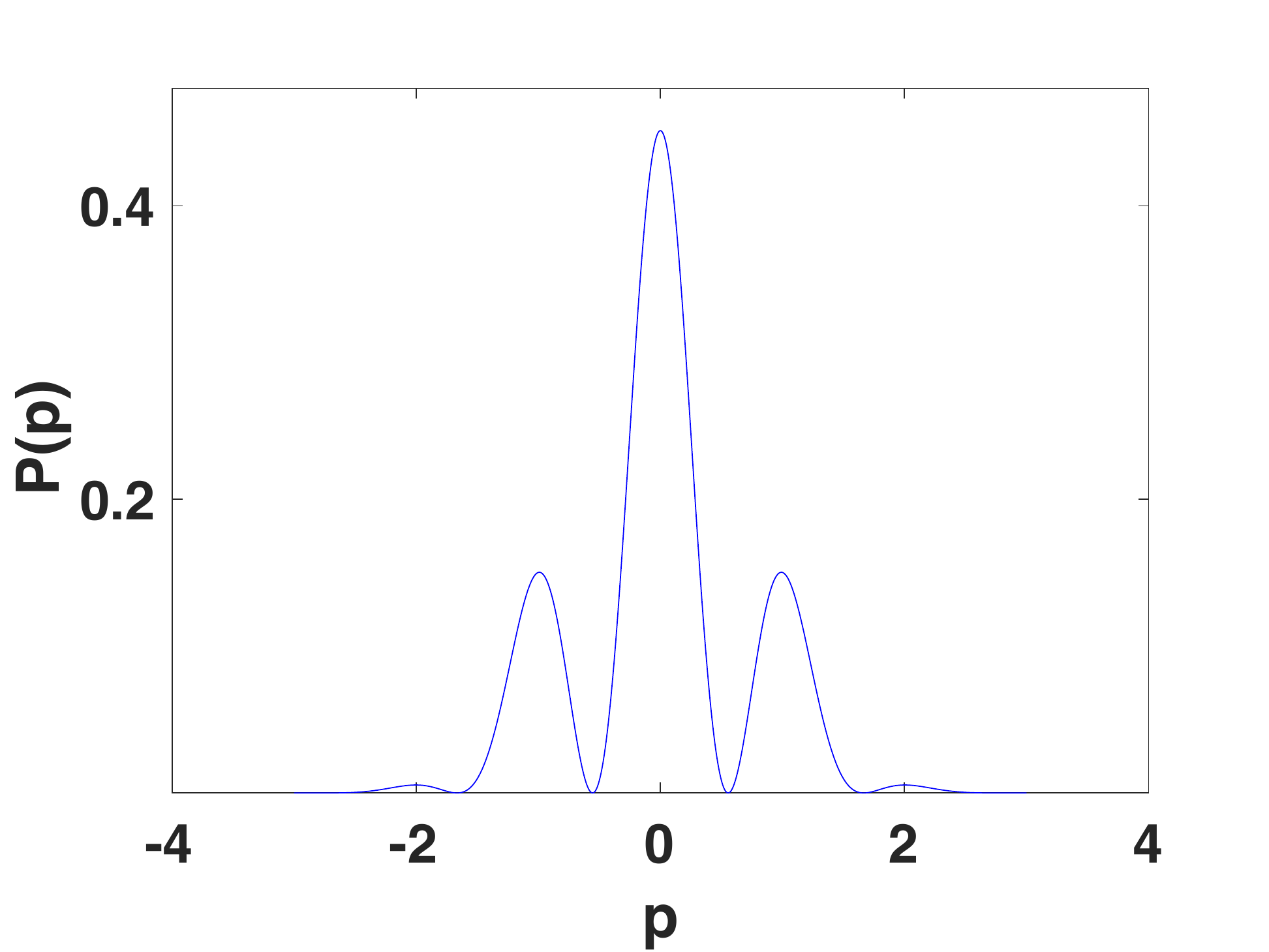}
\par\end{centering}
\caption{\label{fig:Probability-density-distribution-p}Probability density
distributions $P(p)$ for the $p$ quadrature of the cat state Eq.
(\ref{eq:cat_state}) with $\alpha_{0}=2$ as given in Eq. (\ref{eq:P(p)_distribution}).}
\end{figure}
 On the other hand, the $p$ quadrature probability distribution,
$P\left(p\right)$, is given by:

\begin{align}
P\left(p\right) & =\langle p|\hat{\rho}_{cat}|p\rangle\label{eq:P(p)_distribution}\\
 & =\frac{1}{\sqrt{\pi}\mathcal{N}}\left\{ 2exp\left(-p^{2}\right)\left[1+\cos\left(2\sqrt{2}p\alpha_{0}\right)\right]\right\} \,.\nonumber 
\end{align}
This contains a cosine term that gives rise to interference fringes.
As shown in Fig. \ref{fig:Probability-density-distribution-p}, interference
fringes arise in the $p$ quadrature probability distribution for
the cat-state (\ref{eq:cat_state}). In contrast, for a statistical
mixture of two coherent states $|\alpha_{0}\rangle$ and $|-\alpha_{0}\rangle$,
the same quantity will show no fringes. 

\subsection{Wigner function and Wigner negativity}

The Wigner function, introduced by Wigner \citep{PhysRev.40.749},
provides a joint probability distribution $W\left(x,p\right)$ of
any two conjugate variables $x,p$ for a quantum state. Wigner functions
satisfy a set of mathematical properties that one normally associates
with a probability distribution \citep{HILLERY1984121}. This is certainly
true for the marginal distributions. For instance, the marginal distribution
for $x$ is given by
\begin{align}
P\left(x\right) & =\intop_{-\infty}^{\infty}W\left(x,p\right)\,dp\label{eq:wigner_marginal}
\end{align}
as for a probability distribution. However, there exist quantum states
for which the corresponding Wigner function admits negative values.
In this case, the Wigner function cannot be viewed as a probability
distribution, but rather is a quasi-probability distribution. The
negativity is usually attributed to the non-classicality of the corresponding
quantum state.

A cat state is a highly nonclassical physical state that has a Wigner
function which admits negative values. In the following, we derive
the expression for a cat state Wigner function, which can be obtained
from the characteristic function, the Fourier transform of the Wigner
function. In particular, we use the Weyl-ordered characteristic function
$\chi_{0}$:
\begin{align}
\chi_{0}\left(\lambda\right) & =\text{Tr}\left(\hat{\rho}_{cat}e^{\lambda\hat{a}^{\dagger}-\lambda^{*}\hat{a}}\right)\,.\label{eq:weyl_character_function}
\end{align}
Introducing the complex variable $\lambda$, complementary to $\alpha$,
the corresponding Wigner function is then given by
\begin{eqnarray}
W\left(\alpha\right) & = & \intop\text{exp}\left(-\lambda\alpha^{*}+\lambda^{*}\alpha\right)\chi_{0}\left(\lambda\right)\,\frac{d^{2}\lambda}{\pi^{2}}\,,\nonumber \\
\label{Wigner_function}
\end{eqnarray}
where we use $\int..d^{2}\lambda$ to indicate an integral over the
entire complex plane. For the cat-state density operator in Eq. (\ref{eq:cat_state_density_op}),
$\chi_{0}$ consists of four terms and the corresponding Wigner function
can be shown to be 
\begin{eqnarray}
W\left(\alpha\right) & = & \frac{2}{\pi\mathcal{N}}\left\{ exp\left[-2\left(\alpha^{*}-\alpha_{0}^{*}\right)\left(\alpha-\alpha_{0}\right)\right]\right.\nonumber \\
 &  & +exp\left[-2\left(\alpha^{*}+\alpha_{0}^{*}\right)\left(\alpha+\alpha_{0}\right)\right]\nonumber \\
 &  & +\langle\alpha_{0}|-\alpha_{0}\rangle exp\left[-2\left(\alpha^{*}-\alpha_{0}^{*}\right)\left(\alpha+\alpha_{0}\right)\right]\nonumber \\
 &  & \left.+\langle-\alpha_{0}|\alpha_{0}\rangle exp\left[-2\left(\alpha^{*}+\alpha_{0}^{*}\right)\left(\alpha-\alpha_{0}\right)\right]\right\} \,.\nonumber \\
\label{eq:Wigner_cat_state}
\end{eqnarray}

\begin{figure}[H]

\includegraphics[width=0.9\columnwidth]{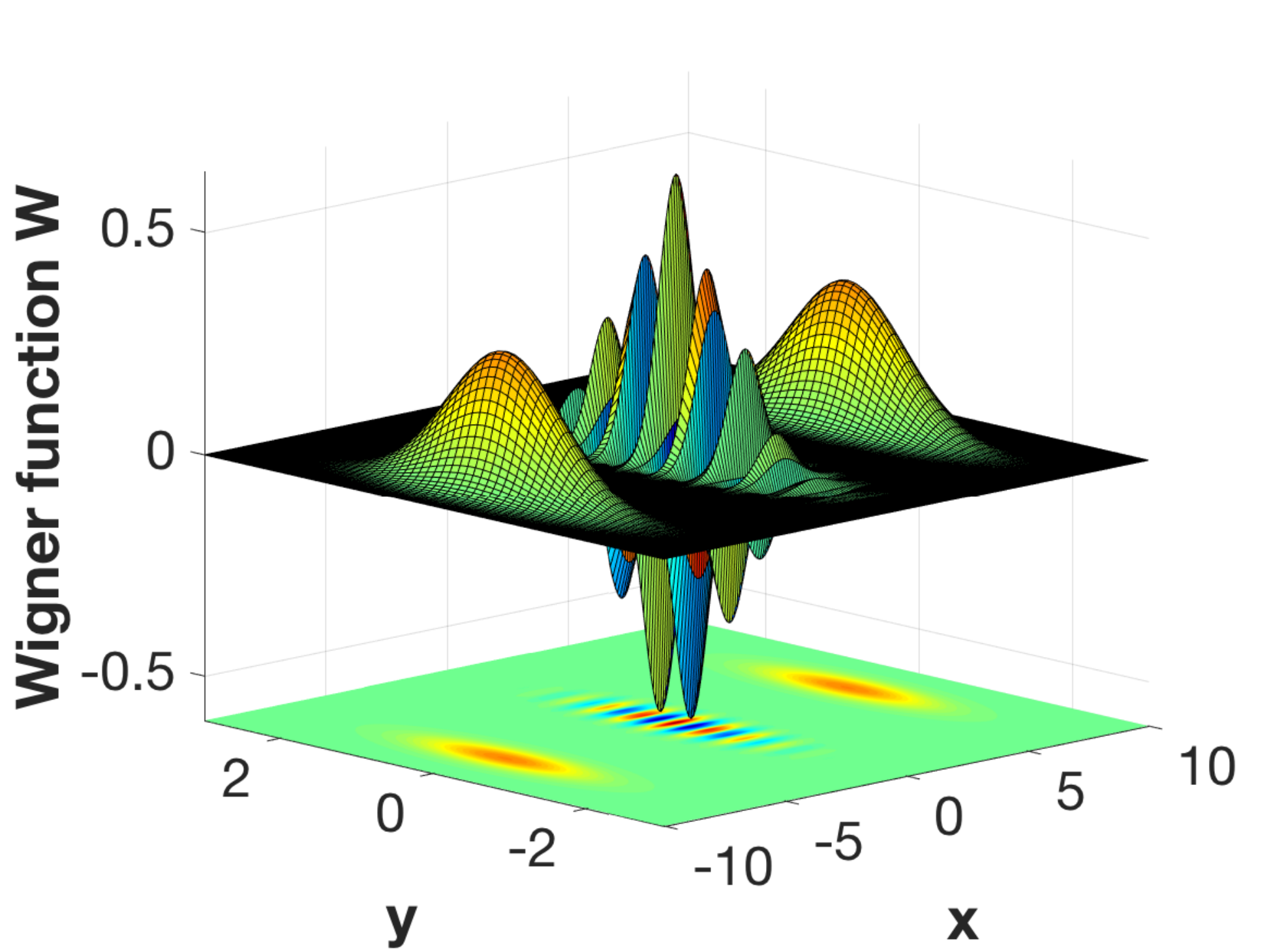}

\caption{\label{fig:cat_wigner}The Wigner function of a cat state as given
in Eq. (\ref{eq:Wigner_cat_state}) for a coherent amplitude $\alpha_{0}=5$.
Here, $x$ and $y$ in the plot are the real and imaginary part of
$\alpha$ in the Wigner function $W\left(\alpha\right)$ respectively.
\textcolor{red}{}}
\end{figure}

The first two terms correspond to the diagonal terms in the cat-state
density operator and are Gaussian distributed, while the last two
terms correspond to the off-diagonal terms in the density operator.
The Wigner function in Eq. (\ref{eq:Wigner_cat_state}) for $\alpha_{0}=5$
is plotted in Fig. \ref{fig:cat_wigner}. In terms of experimental
measurements, the superposition of $\alpha_{0}$ and $-\alpha_{0}$
corresponds to a squared phase-space distance of $S=\left|2\alpha_{0}\right|^{2}=100$,
which has been achieved in superconducting microwave experiments \citep{Vlastakis607}. 

The two Gaussian peaks arise from the first two terms in Eq. (\ref{eq:Wigner_cat_state})
while the region that admits negative values comes from the last two
terms in Eq. (\ref{eq:Wigner_cat_state}). 

The Wigner function has been measured in experiments. For instance,
the Wigner functions of nonclassical photon states in a cavity are
directly measured \citep{PhysRevLett.89.200402} using the Lutterbach
and Davidovich procedure \citep{PhysRevLett.78.2547}. Using the same
procedure, the Wigner function of a two mode cat state is measured
more recently \citep{Wang1087}. We note that these experiments involve
probing the cavity photon state with atoms, which is different from
the quantum memory protocol proposed in this paper. To measure quantum
states of light, homodyne tomography \citep{RevModPhys.81.299} is
needed and this has been carried out both in optomechanical experiments
\citep{Reed:2017aa,Pfaff:2017aa} and in experiments that generate
optical cat states \citep{Sychev:2017aa,Asavanant:17}.

Once we have the Wigner function, we can quantify the negativity of
the Wigner function by introducing the negative volume $\delta$,
which is defined to be \citep{1464-4266-6-10-003}
\begin{eqnarray}
\delta & = & \frac{1}{2}\intop\left[\left|W\left(\alpha\right)\right|-W\left(\alpha\right)\right]\,d^{2}\alpha\,.\label{eq:negative_volume_Wigner}
\end{eqnarray}
A factor of $1/2$ in the definition above means that the Wigner negativity
$\delta$ takes values between $0$ and $1$, and any value larger
than $0$ implies that the Wigner function $W$ has negative values.

\subsection{Reconstruction of the density operator}

We note that the negativity of a Wigner function is not sufficient
to imply the existence of a cat state; it merely signifies the nonclassicality
of the state. We get a clearer picture from the presence or absence
of the off-diagonal terms in the density operator. In principle, a
Wigner function contains all the statistical information about a physical
state and hence a density operator can be obtained from a Wigner function.
This is done in Section \ref{subsec:Numerical-results}. 

In practice, however, a density operator obtained from an experimentally
characterized Wigner function might not be completely positive \citep{PhysRevA.86.032106},
which is unphysical. This is due to the fact that only a finite number
of measurements is recorded in an experiment. Usually, some maximum-likelihood
procedure is carried out to find the most likely density operator
that characterizes a physical state in an experiment \citep{PhysRevA.86.032106,Wang1087}. 

\begin{figure}[H]
\includegraphics[width=0.9\columnwidth]{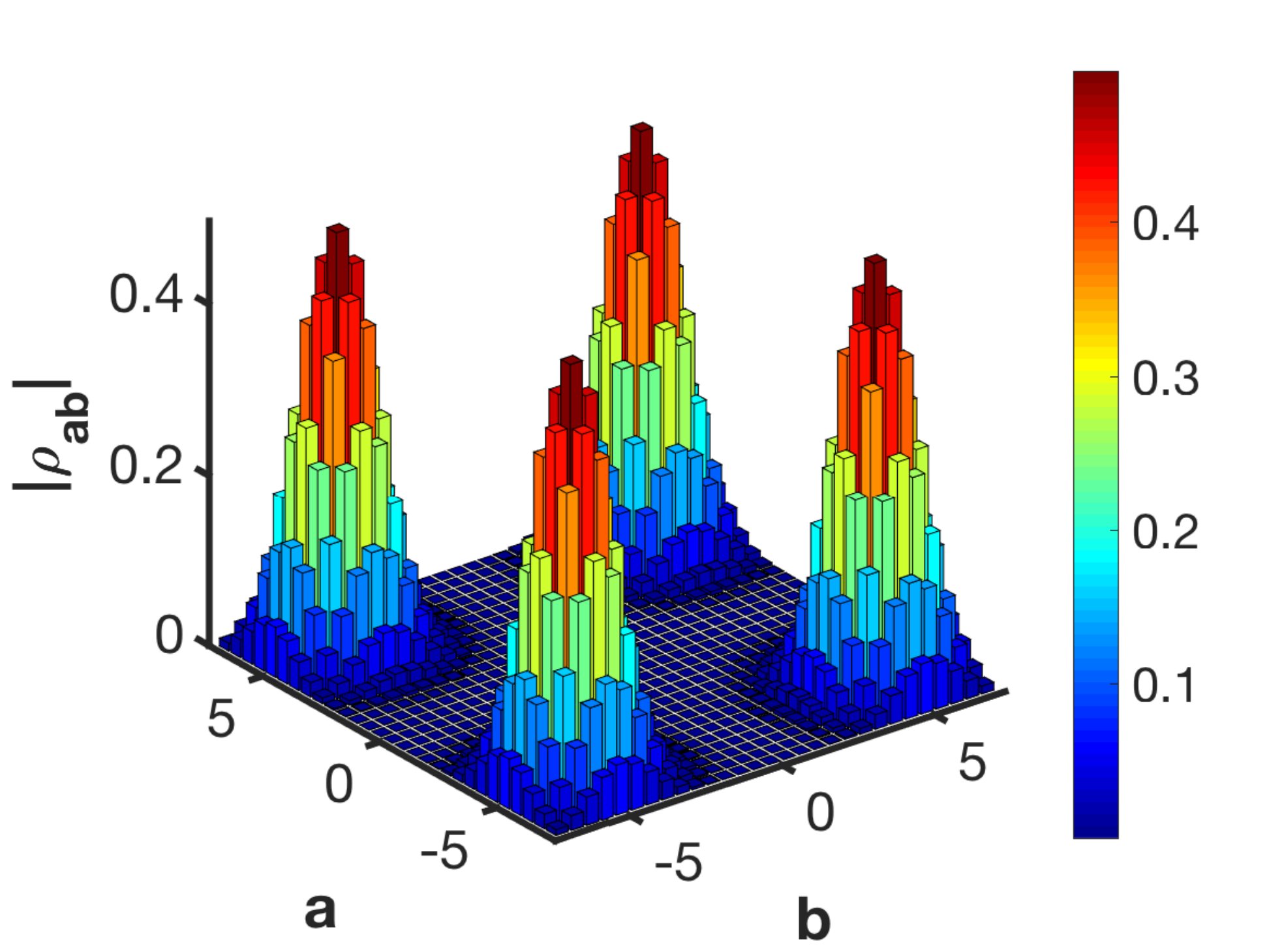}

\caption{The modulus of the density operator for a cat state in the coherent
state basis as given in Eq. (\ref{eq:density_op_reconstruction})
with coherent amplitude $\alpha_{0}=5$ in the coherent state basis
based on Eq. (\ref{eq:density_op_reconstruction}). \label{fig:reconstructed_density_operator}}
\end{figure}

The modulus of the cat state density operator in the coherent state
basis is obtained by
\begin{align}
\left|\langle a|\rho_{cat}|b\rangle\right| & =\frac{1}{\mathcal{N}}\left|\left(\langle a|\alpha_{0}\rangle\langle\alpha_{0}|b\rangle+\langle a|-\alpha_{0}\rangle\langle-\alpha_{0}|b\rangle\right.\right.\nonumber \\
 & \left.\left.+\langle a|\alpha_{0}\rangle\langle-\alpha_{0}|b\rangle+\langle a|-\alpha_{0}\rangle\langle\alpha_{0}|b\rangle\right)\right|\,,\label{eq:density_op_reconstruction}
\end{align}
where $a$, $b$ and $\alpha_{0}$ are taken here to be real for simplicity.
Fig. \ref{fig:reconstructed_density_operator} shows the modulus of
the cat state density operator in the coherent state basis using Eq.
(\ref{eq:density_op_reconstruction}). The presence of off-diagonal
terms implies the quantum superposition between the two distinguishable
coherent states $|\alpha_{0}\rangle$ and $|-\alpha_{0}\rangle$.

\subsection{Variance method}

Alternatively, the cat-state can be distinguished from the mixture
$\hat{\rho}=P_{+}|\alpha_{0}\rangle\langle\alpha_{0}|+P_{-}|-\alpha_{0}\rangle\langle-\alpha_{0}|$
by the method of variances. Variance methods have been used previously
to detect quantum coherences \citep{PhysRevLett.97.170405,PhysRevA.76.030101,PhysRevA.77.062108,1367-2630-14-9-093039,PhysRevA.89.012116,FROWIS20152,Oudot:15,PhysRevLett.116.090801,PhysRevA.94.062125}.
If the system is indeed in a mixture of two states $\hat{\rho}_{+}=|\alpha_{0}\rangle\langle\alpha_{0}|$
and $\hat{\rho}_{-}=|-\alpha_{0}\rangle\langle-\alpha_{0}|$, then
it is straightforward to show that the variance in the $p$ quadrature
must satisfy 
\begin{equation}
(\Delta p)_{mix}^{2}\geq\frac{1}{2}\,.\label{eq:varmix}
\end{equation}
This follows by considering that for any mixture $\hat{\rho}_{mix}=\sum_{R}P_{R}\hat{\rho}_{R}$
of states $\hat{\rho}_{R}$ with probability weightings $P_{R}$,
the variance $(\Delta p)_{mix}^{2}$ of the mixture cannot be less
than the weighted sum of the variances $(\Delta p)_{R}^{2}$ of the
components $\hat{\rho}_{R}$ of the mixture: $(\Delta p)_{mix}^{2}\geq\sum_{R}P_{R}(\Delta p)_{R}^{2}$.
Next we use that for all quantum states $\hat{\rho}_{R}$, $(\Delta x)_{R}(\Delta p)_{R}\geq1/2$,
and that for the coherent states $|\alpha_{0}\rangle$ and $|-\alpha_{0}\rangle$,
it is true that $(\Delta x)_{R}^{2}=\frac{1}{2}$. From this, one
can see that for each component of the mixture $(\Delta p)_{R}^{2}\geq\frac{1}{2}$,
and the result (\ref{eq:varmix}) follows. 

In fact, the result (\ref{eq:varmix}) is predicted for \emph{any}
mixture $\hat{\rho}_{mix}=P_{+}\hat{\rho}_{+}+P_{-}\hat{\rho}_{-}$
for which the variances of $x$ for $\hat{\rho}_{\pm}$ are assumed
to be respectively $(\Delta x)_{\pm}^{2}\leq\frac{1}{2}$. The experimental
observation of $(\Delta p)^{2}<1/2$ in association with the distribution
functions shown in Figure 2 thus confirms that the system is not in
a mixture of any two states $\hat{\rho}_{\pm}$, that each generate
the Gaussian distributions $P_{\pm}(x)\sim exp\left[-\left(x\mp\sqrt{2}\alpha_{0}\right)^{2}\right]$
evident in the $P(x)$. 

Calculation of the variance of $p$ for the cat-state (\ref{eq:cat_state})
gives
\begin{equation}
(\Delta p)_{cat}^{2}=\frac{1}{2}-\frac{2\alpha_{0}^{2}exp\left(-2\alpha_{0}^{2}\right)}{1+exp\left(-2\alpha_{0}^{2}\right)}\label{eq:varcat}
\end{equation}
in clear violation of (\ref{eq:varmix}) for all $\alpha_{0}$. The
observation of $(\Delta p)^{2}<1/2$ is a falsification of the mixed
state $\hat{\rho}_{mix}$. Even for quite small $\alpha_{0}$, this
becomes exceedingly difficult to measure. However, we will see below
that there exist regimes of parameter space where ${\color{black}{\color{black}{\color{black}}}(\Delta p)^{2}<1/2}$
for a non-negative Wigner function.

\section{Phase-space simulations \label{sec:Numerical-simulation-and}}

Having identified Schr\"odinger cat characteristic signatures and
expected properties, we now wish to analyze our more realistic optomechanical
quantum memory model. This has two relevant coupled modes which can
be macroscopically occupied, together with input, output, and reservoir
modes. For this, we turn to a more powerful method: the positive-P
phase-space representation \citep{0305-4470-13-7-018}. This has the
advantage that it can readily treat large, entangled Hilbert spaces,
together with thermal noise, dissipation, and if necessary nonlinear
effects as well \citep{PhysRevA.90.043805,1402-4896-91-7-073007}.

\subsection{Positive-P representation \label{sec:Phase-space-methods}}

The master equation given in Eq. (\ref{eq:master equation}) is an
operator equation and is generally intractable, especially if there
is any nonlinearity. Phase space methods can be used to transform
this operator equation into a set of stochastic differential equations
describing the dynamics of the optical, mechanical and reservoir modes
in an optomechanical system. This is achieved by noting that it is
always possible to represent the quantum density operator $\hat{\rho}$
as an expansion of a positive probability $P\left(\vec{\bm{\alpha}}\right)$
and a set of non-orthogonal projection operators $\hat{\Lambda}\left(\vec{\bm{\alpha}}\right)\,$
\begin{align}
\hat{\rho} & =\intop P\left(\vec{\bm{\alpha}}\right)\hat{\Lambda}\left(\vec{\bm{\alpha}}\right)\,d^{2}\vec{\bm{\alpha}}.\label{eq:general_rho_representation}
\end{align}
In the general case, $\vec{\bm{\alpha}}=\left(\bm{\alpha},\bm{\alpha^{+}}\right)$
is a complex vector consisting of two independent complex vectors
for each mode, namely $\bm{\alpha}=\left(\alpha,\beta,\bm{\alpha}^{in},\bm{\alpha}^{out}\right)$
and $\bm{\alpha^{+}}=\left(\alpha^{+},\beta^{+},\bm{\alpha}^{in+},\bm{\alpha}^{out+}\right)$,
where $\bm{\alpha}$ corresponds to an operator vector $\bm{a}$,
and $\bm{\alpha}^{+}$ corresponds to $\bm{a}^{\dagger}$.

Here $\hat{\Lambda}\left(\vec{\bm{\alpha}}\right)$ is a set of projection
operators parametrized by $\vec{\bm{\alpha}}$ that forms a complete
basis, $P\left(\vec{\bm{\alpha}}\right)$ is the corresponding quasi-probability
density function, and $d^{2}\vec{\bm{\alpha}}$ is an integration
measure over the relevant complex space. There are different ways
that this can be done, depending on the mapping used. In this paper,
the positive-P representation is used, so that the projection operator
$\hat{\Lambda}$ is \citep{0305-4470-13-7-018} 
\begin{align}
\hat{\Lambda}\left(\vec{\bm{\alpha}}\right) & =\frac{|\bm{\alpha}\rangle\langle\bm{\alpha}^{+*}|}{\langle\bm{\alpha}^{+*}|\bm{\alpha}\rangle}=\prod_{m}\hat{\Lambda}\left(\vec{\alpha}_{m}\right)\,,\label{eq:posp_operator_basis}
\end{align}
where $|\bm{\alpha}\rangle$ is a multimode coherent state \citep{Glauber1963_CoherentStates}
and $\bm{\alpha}$ is the corresponding vector coherent state amplitude,
while $\vec{\alpha}_{m}=\left(\alpha_{m},\alpha_{m}^{+}\right)$ gives
the mode amplitude in the $m$-th mode. This approach generalizes
Glauber's P-representation \citep{Glauber_1963_P-Rep}, thus allowing
the inclusion of nonclassical states.

A set of operator identities enables a transformation of the master
equation Eq. (\ref{eq:master equation}) into a Fokker-Planck equation.
A probability distribution with a Fokker-Planck equation having positive-definite
diffusion always exists in the positive-P representation and hence
no truncation approximation is required. The numerical solutions are
then exact, apart from the sampling error which can be arbitrarily
reduced by increasing the number of samples in a simulation. Possible
issues arising from boundary terms, \citep{Gilchrist_Gardiner_PD_PPR_Application_Validity}
which can be otherwise removed \citep{Deuar:2002}, do not appear
here. The positive-P representation has the virtue of always being
positive, even for quantum states that are highly non-classical, as
for instance with cat states. This allows the probabilistic sampling
of quantum states. P-functions of this type have been used previously
to represent cat states generated dynamically in non-equilibrium parametric
oscillators \citep{wolinsky1988quantum,reid1993macroscopic,krippner1994transient}.
Here we assume that the cat state is already generated, and study
how to transfer it to a mechanical oscillator.

\subsection{Stochastic differential equations}

From the Fokker-Planck equation, we obtain a corresponding set of
stochastic differential equations that describe the time evolution
of the cavity $\alpha,\alpha^{+}$ and mechanical $\beta,\beta^{+}$
mode amplitudes. The number of phase space variables is doubled in
the positive-P representation where a mode is characterized by two
phase space variables in order to represent quantum superpositions.
The stochastic differential equations for both the cavity and mechanical
mode amplitudes are given by:
\begin{eqnarray}
d\alpha & = & \left(-\gamma_{o}\alpha-ig(t)\beta\right)dt+d\phi^{in}\nonumber \\
d\alpha^{+} & = & \left(-\gamma_{o}\alpha^{+}+ig(t)\beta^{+}\right)dt+d\phi^{in+}\nonumber \\
d\beta & = & \left(-\gamma_{m}\beta-ig(t)\alpha\right)dt+\sqrt{2\gamma_{m}}d\phi_{m}^{in}\nonumber \\
d\beta^{+} & = & \left(-\gamma_{m}\beta^{+}+ig(t)\alpha^{+}\right)dt+\sqrt{2\gamma_{m}}d\phi_{m}^{in+}\,,\label{eq:sde_posp}
\end{eqnarray}
where $\bm{\alpha},\bm{\alpha^{+}}$ are conjugate in the mean, but
not for individual realizations, and
\begin{eqnarray}
d\phi^{in} & = & \sqrt{2\gamma_{ext}}d\phi_{ext}^{in}+\sqrt{2\gamma_{int}}d\phi_{int}^{in}\nonumber \\
d\phi_{in}^{+} & = & \sqrt{2\gamma_{ext}}d\left(\phi_{ext}^{in}\right)^{+}+\sqrt{2\gamma_{int}}d\left(\phi_{int}^{in}\right)^{+}\,.\label{eq:inputs_posp}
\end{eqnarray}
The terms $\phi_{ext}^{in},\phi_{ext}^{in+}$ are obtained from a
mode expansion in terms of external amplitudes $\bm{\alpha}^{in},\bm{\alpha}^{in+}$,
as in the operator mode expansion, Eq (\ref{eq:Externalfieldmodes}),
so that:
\begin{align}
\phi_{ext}^{in}\left(t\right) & =\sum_{n\ge0}\alpha_{n}^{in}u_{n}^{in}\left(t\right)\label{eq:Externalfieldmodes-1}
\end{align}
 The conjugate terms are obtained by the usual mapping of $\phi\rightarrow\phi^{+}$,
$\alpha\rightarrow\alpha^{+}$ and $u_{n}\rightarrow u_{n}^{*}$.
However, $\phi_{m}^{in},\phi_{int}^{in},\phi_{m}^{in+},\phi_{int}^{in+}$
are Langevin noise terms obtained from transforming the master equation
(\ref{eq:master equation}) into a Fokker-Planck equation, using the
standard positive-P identities \citep{0305-4470-13-7-018}. 

The effective optomechanical coupling strength $g\left(t\right)$
is time dependent due to the optomechanical state transfer protocol
used. It is a constant during the writing and readout stages, and
zero during the storing stage: 
\begin{align}
g\left(t\right)= & \begin{cases}
\sqrt{N}g_{0}, & -t_{w}\leq t\leq0\\
0, & 0\leq t\leq t_{s}\quad,\\
\sqrt{N}g_{0}, & t_{s}\leq t\leq t_{r}
\end{cases}\label{eq:time_dependent_g}
\end{align}
where $t_{w}$, $t_{s}$, and $t_{r}=t_{w}$ are the durations for
the writing, storing, and readout stages, respectively, and $N$ is
the intra-cavity pump photon number.

The external cavity input $\phi_{ext}^{in},\phi_{ext}^{in+}$ contain
the information about the cat state to be stored in the mode amplitude
$\alpha_{0}^{in},\alpha_{0}^{in+}$. Apart from this, the other input
modes are assumed to be in vacuum states. The internal cavity $\phi_{int}^{in},\phi_{int}^{in+}$,
and mechanical $\phi_{m}^{in},\phi_{m}^{in+}$ inputs are in thermal
equilibrium, and satisfy the following normally ordered correlations:
\begin{eqnarray}
\langle d\phi_{i}^{in}d\phi_{j}^{in+}\rangle & = & \bar{n}_{i,th}\delta_{ij}dt\,,\label{eq:posp_thermal_correlation}
\end{eqnarray}
where the indices $i,j=1,2\sim int,m$, and $\bar{n}_{i,th}$ are
the mean thermal occupations. In this work, the experimental parameters
used mean that only the mechanical thermal bath contributes significantly.
The optical thermal noises are neglected. 

The input mode into the cavity and output mode from the cavity are
related by the input-output relation $\phi_{ext}^{out}\left(t\right)=\sqrt{2\gamma_{ext}}\alpha\left(t\right)-\phi_{ext}^{in}\left(t\right)$
\citep{PhysRevA.31.3761}, together with a conjugate equation. The
integrated output $\alpha_{0}^{out},\alpha_{0}^{out+}$ mode amplitudes
can be obtained by integrating these modes with temporal mode functions
$u_{0}^{in}\left(t\right)$ and $u_{0}^{out}\left(t\right)$ as given
below:
\begin{eqnarray}
\alpha_{0}^{out} & = & \intop_{t_{s}}^{\infty}u_{0}^{out}\left(t\right)\phi_{ext}^{out}\left(t\right)\,dt\,,\label{eq:integrated_input_output}
\end{eqnarray}
where $u_{0}^{out}\left(t\right)$ is given by a time reversed version
of Eq. (\ref{eq:input_mode_function}), defined for $t>t_{s}$, and
$t_{s}$ is the storage time. The integrated output mode amplitudes
$\alpha_{0}^{out+}$ are defined similarly. 

The input mode function $u_{0}^{in}$ in Eq. (\ref{eq:input_mode_function})
has the form $\left[e^{\left(\gamma_{+}+m\right)t}-e^{\left(\gamma_{+}-m\right)t}\right]\Theta\left(-t\right)$,
and it can be shown that \citep{PhysRevA.96.013854} in the limit
where $\gamma_{m}\ll g\ll\gamma_{o}$, then $e^{\left(\gamma_{+}-m\right)t}\Theta\left(-t\right)$
is the dominating term during the writing stage. This suggests that
the duration of the writing stage has to be longer than $1/\left(\gamma_{+}-m\right)$.
However, we use the exact mode-function in our calculations. In this
work, we choose the writing stage duration to be $10/\left(\gamma_{+}-m\right)$.
The storage time $t_{s}$ is chosen to be some fraction of the mechanical
lifetime. Finally, the read-out stage has the same duration as the
writing stage.

\subsection{Cat state and importance sampling}

Initially, we assume that only the external cat state in mode $a_{0}^{in}$
is excited, so that 
\begin{equation}
\hat{\rho}=\hat{\rho}_{cat}\otimes\hat{\rho}',
\end{equation}
where $\hat{\rho}_{cat}$ is the state of the input mode $a_{0}^{in}$,
and $\hat{\rho}'$ is the state of all the remaining modes, which
are assumed to be in the vacuum state, except that the mechanical
mode may be initially thermally excited. The cat density operator
$\hat{\rho}_{cat}$ in Eq. (\ref{eq:cat_state_density_op}) can be
expressed in the positive-P representation as follows:
\begin{align}
\hat{\rho}_{cat} & =\intop\intop P\left(\vec{\alpha}_{0}^{in}\right)\hat{\Lambda}\left(\vec{\alpha}_{0}^{in}\right)\,d^{2}\vec{\alpha}_{0}^{in}\,.\label{eq:pos_p_rep}
\end{align}
One of the possible compact positive-P distributions for the cat state
Eq. (\ref{eq:cat_state}) is given by \citep{0305-4470-13-7-018,1402-4896-91-7-073007}
\begin{align}
P\left(\vec{\alpha}_{0}^{in}\right) & =\frac{1}{\mathcal{N}}\left[\delta_{+,+}+\delta_{-,-}+e^{-2\left|\alpha_{0}\right|^{2}}\left(\delta_{+,-}+\delta_{-,+}\right)\right]\,,\nonumber \\
\label{eq:posP_dist_cat}
\end{align}
where $\delta_{\pm,\pm}=\delta\left(\alpha_{0}^{in}\pm\alpha_{0}\right)\delta\left(\alpha_{0}^{in+*}\pm\alpha_{0}\right)$.
It is straightforward to show that the positive-P distribution in
Eq. (\ref{eq:posP_dist_cat}) gives the correct density operator in
Eq. (\ref{eq:cat_state_density_op}). This distribution is particularly
easy to sample. One draws a sample of $\alpha_{0}^{in}$ and $\alpha_{0}^{in+}$
with values from one of the possible four terms with the corresponding
probability as given in Eq. (\ref{eq:posP_dist_cat}). 

In order to carry out positive-P simulations, an ensemble of input
coherent amplitudes $\alpha_{0}^{in}$ and $\alpha_{0}^{in+}$ that
corresponds to the correct cat-state statistics has to be sampled
from the positive-P distribution in Eq. (\ref{eq:posP_dist_cat}).
In particular, the last two terms in Eq. (\ref{eq:posP_dist_cat})
arise from the off-diagonal terms in the cat-state density operator
which is the source of non-classicality in a cat-state. 

For the case where $\alpha_{0}$ is large, the off-diagonal events
are rare in samples taken from the standard positive-P distribution.
However, they can have a large effect on some observables. The task
is to include these rare, but significant terms in our samples. This
is achieved using the importance sampling method, whereby a different
distribution is used such that these rare terms are sampled sufficiently.
When doing this, both the kernel function $\hat{\Lambda}$ and the
probability distribution are modified so as to leave the density operator
invariant. 

The weighted phase space representation of the input mode density
operator is now:
\begin{align}
\hat{\rho}_{cat} & =\intop\intop f\left(\vec{\alpha}_{0}^{in}\right)\hat{\Lambda}_{w}\left(\vec{\alpha}_{0}^{in}\right)\,d^{2}\vec{\alpha}_{0}^{in}\,,\label{eq:importance_sampling_density_op}
\end{align}
where $\hat{\Lambda}_{w}\left(\vec{\alpha}_{0}^{in}\right)\equiv\hat{\Lambda}\left(\vec{\alpha}_{0}^{in}\right)w\left(\vec{\alpha}_{0}^{in}\right)$
is the weighted kernel function with weight $w\left(\vec{\alpha}_{0}^{in}\right)=P\left(\vec{\alpha}_{0}^{in}\right)/f\left(\vec{\alpha}_{0}^{in}\right)$,
associated with the sampling of the distribution $f\left(\vec{\alpha}_{0}^{in}\right)$.
A natural initial distribution choice is a probability distribution
of the form $f\left(\vec{\alpha}_{0}^{in}\right)=\frac{1}{4}\left(\delta_{+,+}+\delta_{-,-}+\delta_{+,-}+\delta_{-,+}\right)$,
with equal probability assigned to each term. Instead of representing
the cat-state density operator $\hat{\rho}_{cat}$ in terms of projection
operators $|\alpha_{0}^{in}\rangle\langle\alpha_{0}^{in+*}|/\langle\alpha_{0}^{in+*}|\alpha_{0}^{in}\rangle$
with the corresponding probability distribution $P\left(\vec{\alpha}_{0}^{in}\right)$,
it is now expressed in terms of an operator $\hat{\Lambda}_{w}\left(\vec{\alpha}_{0}^{in}\right)$,
with the new probability distribution $f\left(\vec{\alpha}_{0}^{in}\right)$.
This weight function has to be taken into account when we compute
any observables. 

The \emph{total} initial density operator can now be written as:
\begin{equation}
\hat{\rho}_{0}=\intop F_{0}\left(\vec{\bm{\alpha}}\right)\hat{\Lambda}_{w}\left(\vec{\bm{\alpha}}\right)\,d^{2}\vec{\bm{\alpha}}\,.
\end{equation}
Here $\hat{\Lambda}_{w}\left(\vec{\bm{\alpha}}\right)\equiv\hat{\Lambda}\left(\vec{\bm{\alpha}}\right)w\left(\vec{\alpha}_{0}^{in}\right)$
and $F_{0}\left(\vec{\bm{\alpha}}\right)=f\left(\vec{\alpha}_{0}^{in}\right)P'\left(\vec{\bm{\alpha}}'\right)$,
where $\vec{\bm{\alpha}}'$ represents the other modes of the system,
initially in a vacuum or thermal state described by the distribution
$P'\left(\vec{\bm{\alpha}}'\right)$. With this new quasi-probability
distribution, $F_{0}\left(\vec{\bm{\alpha}}\right)$, any moments
we compute have to be weighted according to $w\left(\vec{\alpha}_{0}^{in}\right)$
to obtain correct results. This is because $\hat{\Lambda}_{w}\left(\vec{\bm{\alpha}}\right)$
no longer has a unit trace, and in fact for any trace that includes
the weighted input mode, 
\begin{equation}
tr\left[\hat{\Lambda}_{w}\left(\vec{\bm{\alpha}}\right)\right]=w\left(\vec{\alpha}_{0}^{in}\right).
\end{equation}

We also note that, somewhat counter-intuitively, the input mode amplitudes
$\vec{\alpha}_{0}^{in}$ are time-invariant. This is because, in simple
terms, they have a 'use-by' time. The effect of these mode amplitudes
is transmitted to the cavity through the associated time-dependent
mode-function $u_{0}\left(t\right)$, rather than through any change
in the input amplitudes themselves.

\subsection{Wigner function and interference fringes}

In this subsection, we describe how a cat signature can be computed
numerically. The simplest cat signature is an interference fringe,
obtained from homodyne measurements on the output field. This is directly
computable from the density operator, and hence one can obtain a sampled
representation of interference by summing over the stochastic trajectories.
We note that the total density operator $\hat{\rho}$ is a multimode
operator, while the cat signatures are inferred only from the integrated
output modes. To this end, we define a projection operator $|p\rangle\langle p|$
that only acts on the chosen output mode. To evaluate this, it is
simple to trace over the non-observed modes, thus generating a single-mode
density matrix, now defined in terms of the output mode amplitudes
$\vec{\alpha}_{0}^{out}$ . These amplitudes are evaluated through
the integrals of Eq (\ref{eq:integrated_input_output}).

We define the output single-mode density matrix as a partial trace
of the density matrix over all modes \emph{except }the mode-matched
output mode, at the final evolution time of the density matrix:
\begin{equation}
\hat{\rho}_{out}=tr_{\mathcal{H}'}\left[\hat{\rho}(t=t_{f})\right]\,.
\end{equation}
This has a phase-space representation of:
\begin{equation}
\hat{\rho}_{out}=\int P\left(\vec{\alpha}_{0}^{out}|\vec{\alpha}_{0}^{in}\right)w\left(\vec{\alpha}_{0}^{in}\right)\hat{\Lambda}\left(\vec{\alpha}_{0}^{out}\right)d\vec{\alpha}_{0}^{out}d\vec{\alpha}_{0}^{in}.
\end{equation}

Here, $P\left(\vec{\alpha}_{0}^{out}|\vec{\alpha}_{0}^{in}\right)$
is the conditional probability of observing $\vec{\alpha}_{0}^{out}$
given an input amplitude $\vec{\alpha}_{0}^{in}$, and it is obtained
by integrating the P-distribution over all the unobserved modes except
the input and output modes. The output quadrature probability distribution
 can then be computed as follows:
\begin{align}
P\left(p\right) & =\text{Tr}\left[\hat{\rho}_{out}|p\rangle\langle p|\right]\nonumber \\
 & =\intop P\left(\vec{\alpha}_{0}^{out}|\vec{\alpha}_{0}^{in}\right)w\left(\vec{\alpha}_{0}^{in}\right)\text{Tr}\left(\hat{\Lambda}\left(\vec{\alpha}_{0}^{out}\right)\,|p\rangle\langle p|\right)\,d\vec{\alpha}_{0}^{out}d\vec{\alpha}_{0}^{in}\label{prob_dist_pos_p}
\end{align}
The output mode is traced out in the second line of Eq. (\ref{prob_dist_pos_p}).

We compute the probability distribution $P\left(p\right)$ of the
integrated output modes $\alpha_{0}^{out},\,\alpha_{0}^{out+}$ to
verify the presence of cat state in the quantum memory. In the Monte
Carlo method, $P\left(p\right)$ in Eq. (\ref{prob_dist_pos_p}) is
estimated from $N_{s}$ phase-space samples, $\left[\vec{\bm{\alpha}}_{1},\ldots\vec{\bm{\alpha}}_{N_{s}}\right]$.
This is shown explicitly below:
\begin{align}
P\left(p\right) & \approx\frac{1}{N_{s}}\sum_{i=1}^{N_{s}}w\left(\vec{\alpha}_{0,i}^{in}\right)\frac{\langle p|\alpha_{0,i}^{out}\rangle\langle\alpha_{0,i}^{out+}{}^{*}|p\rangle}{\langle\alpha_{0,i}^{out+}{}^{*}|\alpha_{0,i}^{out}\rangle}\,.\label{eq:prob_dis_monte_carlo}
\end{align}
 In particular, samples with index $i$ going from $1$ to $N_{s}/2$
correspond to diagonal terms in the density operator and they have
a weight function $w=2/\left(1+e^{-2\left|\alpha_{0}\right|^{2}}\right)$,
while samples with index $i$ going from $N_{s}/2+1$ to $N_{s}$
correspond to off-diagonal terms in the density operator and the weight
function is $w=2e^{-2\left|\alpha_{0}\right|^{2}}/\left(1+e^{-2\left|\alpha_{0}\right|^{2}}\right)$.
For cases where the mechanical thermal noise $\bar{n}_{th}\neq0$,
the accuracy of the estimation improves with the number of samples
$N_{s}$. At zero temperature there is no sampling error, giving an
extremely efficient procedure.

In order to obtain the Wigner function of the integrated output modes,
it is necessary to relate the positive-P function to its corresponding
Wigner function. We write down the expression of the Wigner function
in terms of the symmetrical-ordered characteristic function and then
represent the density operator $\hat{\rho}_{out}$ in that characteristic
function in the positive-P representation. These steps are explicitly
shown below:

\begin{align}
W\left(\alpha\right) & =\frac{1}{\pi^{2}}\intop\text{e}^{\left(-\lambda\alpha^{*}+\lambda^{*}\alpha\right)}\chi_{W}\left(\lambda\right)\,d^{2}\lambda\nonumber \\
 & =\frac{2}{\pi}\intop P\left(\vec{\alpha}_{0}^{out}|\vec{\alpha}_{0}^{in}\right)w\left(\vec{\alpha}_{0}^{in}\right)\text{e}^{\left[-2\left(\alpha_{0}^{out+}-\alpha^{*}\right)\left(\alpha_{0}^{out}-\alpha\right)\right]}d\vec{\alpha}_{0}^{out}d\vec{\alpha}_{0}^{in}\,\,.\label{eq:wigner_from_pos_p}
\end{align}

In going from line 1 to line 2 in Eq. (\ref{eq:wigner_from_pos_p}),
the characteristic function $\chi_{W}\left(\lambda\right)=\text{Tr}\left(\hat{\rho}_{0}^{out}e^{\lambda\hat{a}^{\dagger}-\lambda^{*}\hat{a}}\right)$
is used, and the density operator $\hat{\rho}_{out}$ is expressed
in the positive-P representation as previously mentioned. Eq. (\ref{eq:wigner_from_pos_p})
is then computed numerically for the integrated output modes $\alpha_{0}^{out},\alpha_{0}^{out+}$
using the Monte Carlo method, giving:
\begin{align}
W\left(\alpha\right)\approx & \frac{2}{\pi N_{s}}\sum_{i}^{N_{s}}w\left(\vec{\alpha}_{0,i}^{in}\right)\text{e }^{\left[-2\left(\alpha_{0,i}^{out+}-\alpha^{*}\right)\left(\alpha_{0,i}^{out}-\alpha\right)\right]}\,.\label{eq:wigner_monte_carlo}
\end{align}
Here, the weight function $w$ is identical to the one given in Eq.
(\ref{eq:prob_dis_monte_carlo}). 

\section{Numerical results \label{subsec:Numerical-results}}

In this section, we describe the numerical method and results for
the cat-state signatures discussed in Section \ref{sec:cat_state_signatures}.
In all of the simulations carried out, both the cavity and mechanical
modes are initially in their ground or thermally excited states. The
cat state is then sent into the cavity, where the cat state is sampled
using the importance sampling method discussed in the previous section.
We generate four different types of positive-P trajectories which
correspond to two diagonal terms and two off-diagonal terms in the
cat state density operator. All numerical simulations were carried
out in the positive-P representation. 

\subsection{Parameter values}

Going through the quantum memory protocol as described in Section
\ref{subsec:Optomechanical-state-transfer}, the output from the cavity
is subsequently integrated to give the output mode amplitudes $\vec{\alpha}_{0}^{out}=\left(\alpha_{0}^{out},\,\alpha_{0}^{out+}\right)$
in Eq. (\ref{eq:integrated_input_output}). These output modes are
the quantum states stored in the quantum memory and all cat state
signatures computed in this section are based on these output modes
amplitudes.

For definiteness, we use experimental parameters from the electromechanical
experiment of Palomaki et al. \citep{palomaki2013coherent}. In their
experiment, the resonator and mechanical decay rates are $\gamma_{o}/2\pi=170\text{kHz}$
and $\gamma_{m}/2\pi=17.5\text{Hz}$ respectively, and the bare electromechanical
coupling strength $g_{0}$ is $2\pi\times200\text{Hz}$.

All numerical simulations are carried out using xSPDE, which is a
Matlab open software package designed specially for solving stochastic
differential equations \citep{Kiesewetter201612}. The algorithm used
for solving the stochastic differential equations is the fourth-order
Runge-Kutta method in the interaction picture \citep{DRUMMOND1991144,Kiesewetter201612}.
As the linearized optomechanical Hamiltonian is used for this work,
the highest frequency parameter in the stochastic differential equations
is the decay rate $\gamma_{o}$. Based on the Shannon sampling theorem
\citep{1697831}, we choose a time step, $\Delta t=1/\left(10\gamma_{o}\right)$
that is smaller than the sufficient sampling rate criterion, which
predicts that a time step less than $1/\left(2\gamma_{o}\right)$
is needed. 

We express all stochastic differential equations in dimensionless
form, using a dimensionless time variable $\tau=\gamma_{o}t$ where
$\gamma_{o}$ is the resonator decay rate. All parameters then have
values that are relative to the resonator decay rate $\gamma_{o}$.
These dimensionless parameters are denoted by capitalizing the Greek
letters of their corresponding experimental parameters. We choose
the dimensionless effective optomechanical coupling strength $G=g/\gamma_{o}=0.6$.
This places the optomechanical system in the weak coupling regime,
where the linearization approximation is valid \citep{PhysRevA.96.013854}.
We take the initial optical and mechanical states to be in their ground
state, except in the last case treated. In simulations where the mechanical
thermal noise $\bar{n}_{th}=0$, we take a total of four samples,
which corresponds to four different trajectories for two diagonal
and two off-diagonal terms in the density operator. In cases where
$\bar{n}_{th}\neq0$, a total number of $2\times10^{5}$ samples are
taken. 

\subsection{Interference fringes }

\begin{figure}[H]

\begin{centering}
\includegraphics[width=0.8\columnwidth]{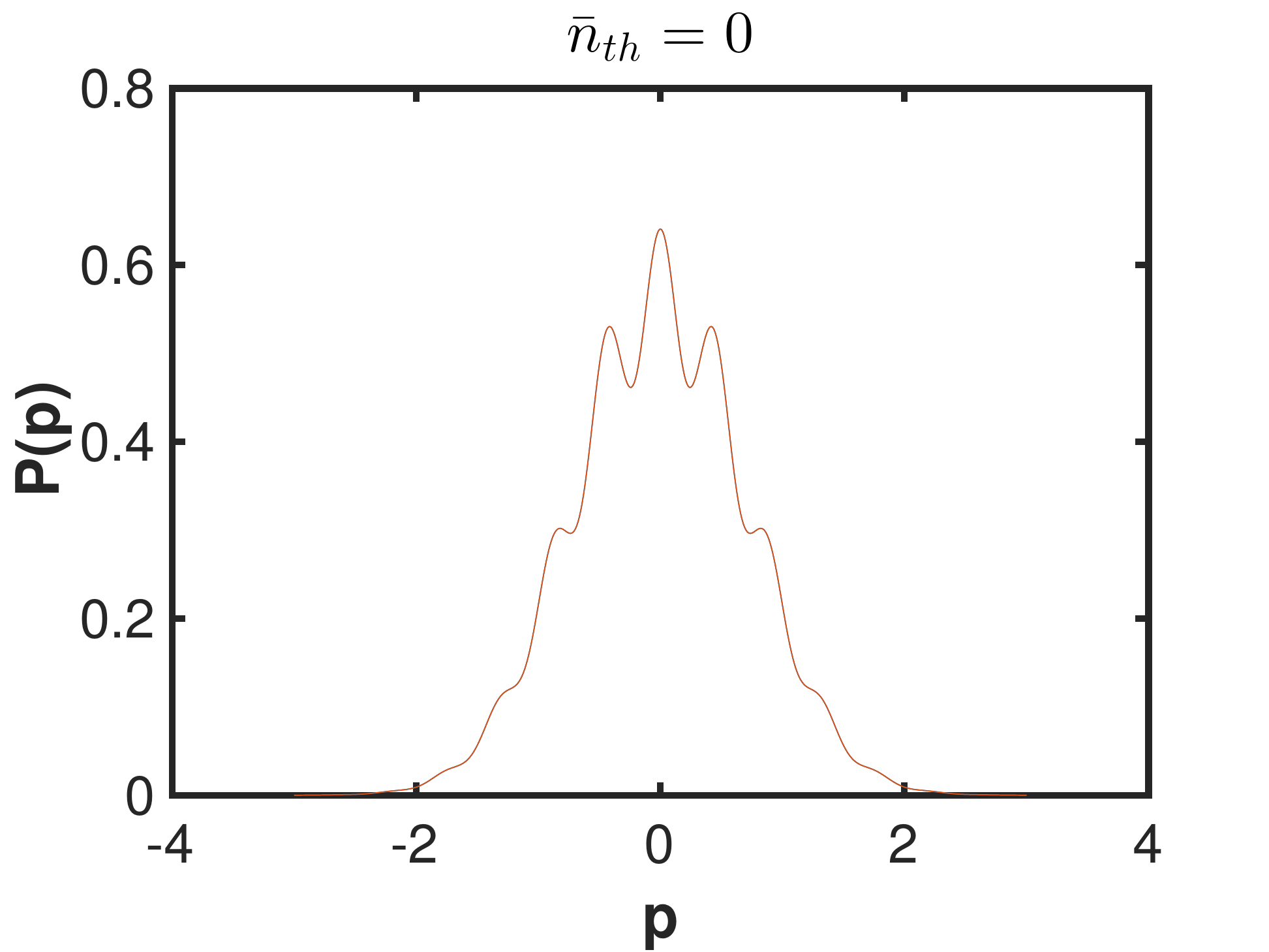}
\par\end{centering}
\caption{The $p$-quadrature probability distribution computed using the positive-P
distribution with Eq. (\ref{eq:prob_dis_monte_carlo}) for $\alpha_{0}=5$
after reading out from the quantum memory. The mean mechanical thermal
noise and internal loss rate are chosen to be $\bar{n}_{th}=\Gamma_{int}=0$
throughout the simulation. Here, the optomechanical cat state has
a low decoherence due to the short storage time compared to the mechanical
oscillator lifetime. \textcolor{black}{This figure is for a storage
time of $0.02/\Gamma_{m}$. }A total number of $4$ samples are taken.
\textcolor{black}{\label{fig:fringes_plot_upperbound-2}}}
\end{figure}

Using the method of Eq (\ref{eq:prob_dis_monte_carlo}) , fringes
were calculated for a cat state with amplitude $\alpha_{0}=5$ corresponding
to $25$ stored phonons. In Fig. \ref{fig:fringes_plot_upperbound-2},
we plot the $p$-quadrature distribution after the readout from the
optomechanical quantum memory. In this figure, there is no internal
cavity loss and the storage time is $0.02/\Gamma_{m}$. 

\begin{figure}[H]

\begin{centering}
\includegraphics[width=0.8\columnwidth]{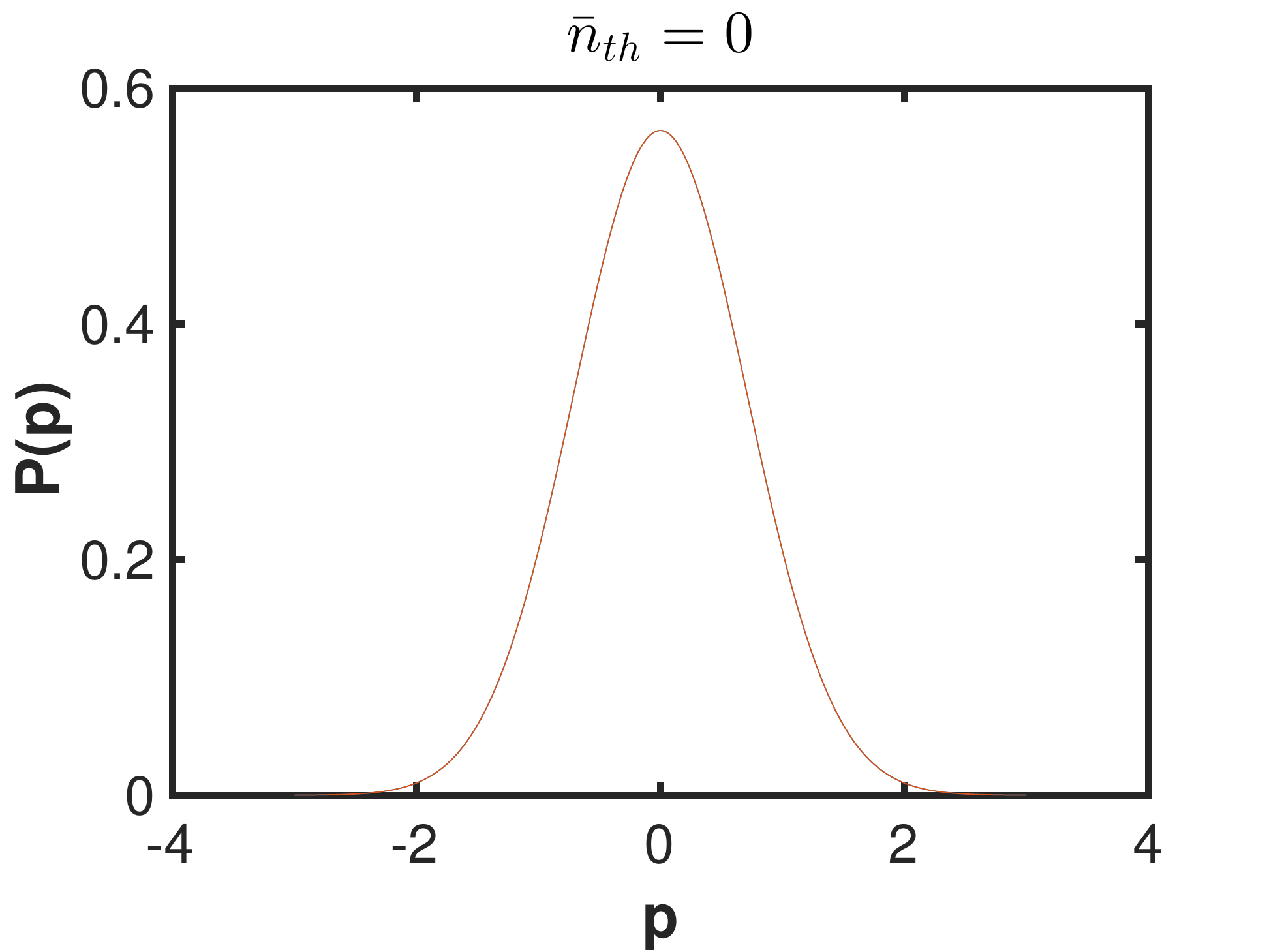}
\par\end{centering}
\caption{The $p$-quadrature probability distribution computed using using
the positive-P distribution with Eq. (\ref{eq:prob_dis_monte_carlo})
for $\alpha_{0}=5$. Other parameters as in Fig (\ref{fig:fringes_plot_upperbound-2}).
Here, the optomechanical cat state decoheres\textcolor{black}{{} for
a storage time of $0.3466/\Gamma_{m}$, which is the time a Wigner
function loses its negativity according to Eq. (\ref{eq:upper_t_wigner_positive}).
\label{fig:fringes_plot_upperbound}}}
\end{figure}

The same quantity but with a storage time $0.3466/\Gamma_{m}$ is
shown in Fig. \ref{fig:fringes_plot_upperbound}. This storage time
corresponds to the time a Wigner function loses its negativity for
a mean mechanical thermal number $\bar{n}_{th}=0$ as given by Eq.
(\ref{eq:upper_t_wigner_positive}). We note in this case, the fringe
pattern has vanished, consistent with a loss of non-classicality.

\subsection{Wigner function }

\begin{figure}[H]

\includegraphics[width=0.9\columnwidth]{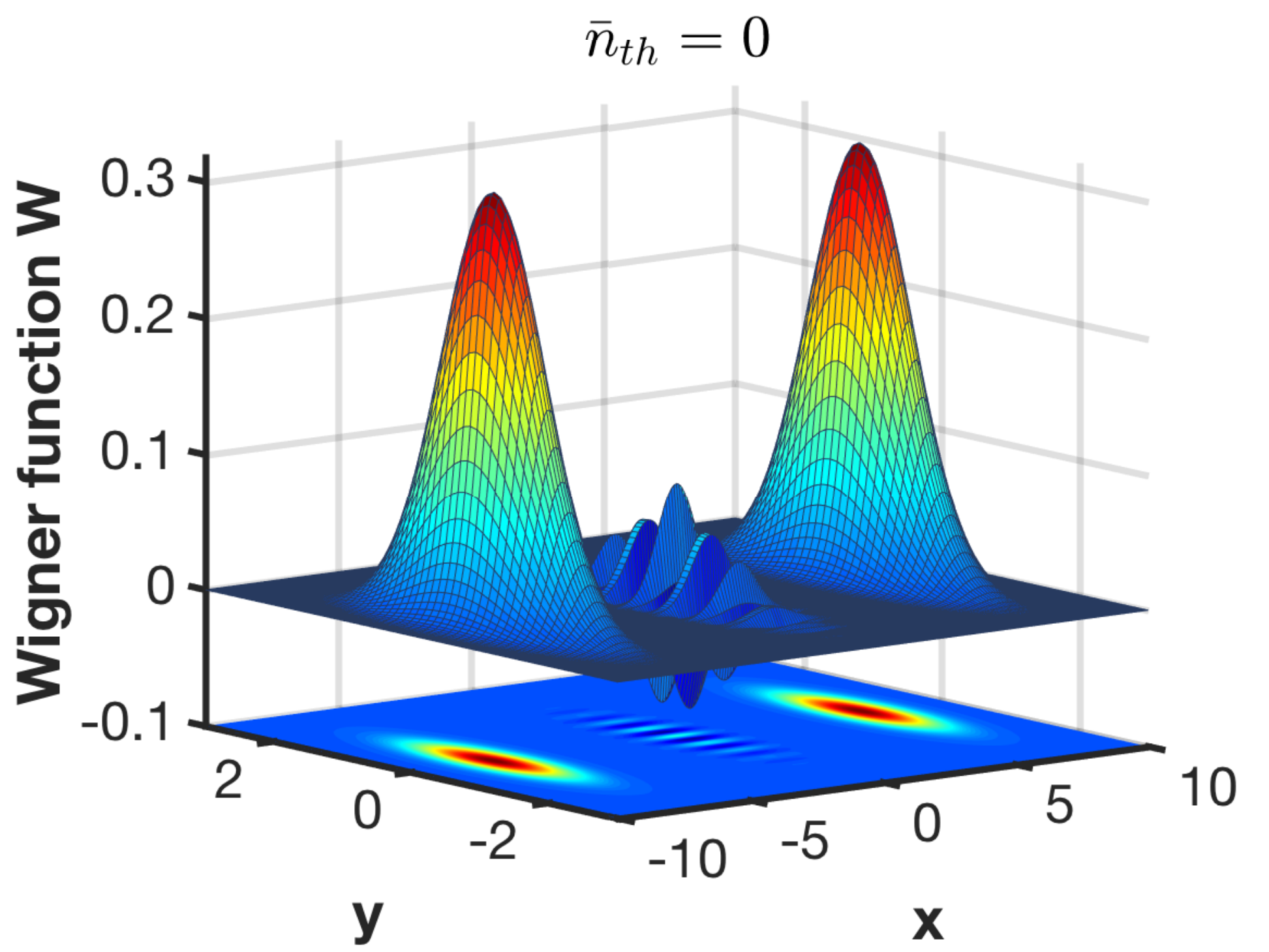}

\caption{\textcolor{red}{}The Wigner function computed using the positive-P
distribution and Eq. (\ref{eq:wigner_monte_carlo}) for $\alpha_{0}=5$
after reading out from the quantum memory. Here, $x$ and $y$ in
the plot are the real and imaginary part of $\alpha$ in the Wigner
function $W\left(\alpha\right)$ in Eq. (\ref{eq:wigner_monte_carlo})
respectively. Other parameters as in Fig (\ref{fig:fringes_plot_upperbound-2}).
This\textcolor{black}{{} figure has a storage time of $0.02/\Gamma_{m}$,
too short for substantial decoherence.  \label{fig:wigner_upperbound-1}}}
\end{figure}

\begin{figure}[H]

\includegraphics[width=0.9\columnwidth]{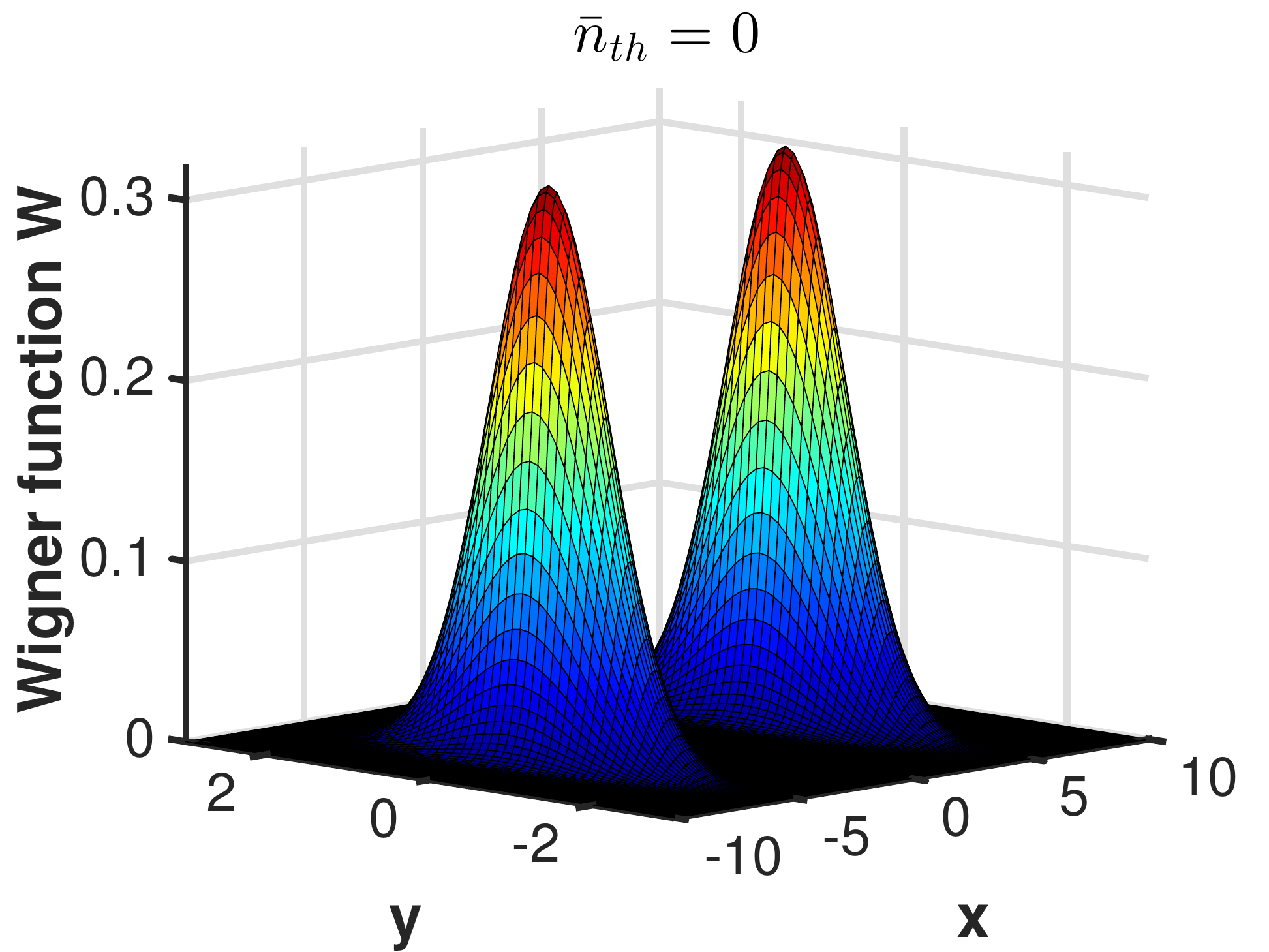}

\caption{\textcolor{red}{}The Wigner function computed using the positive-P
distribution and Eq. (\ref{eq:wigner_monte_carlo}) for $\alpha_{0}=5$
after reading out from the quantum memory. As previously, $x$ and
$y$ in the plot are the real and imaginary part of $\alpha$ in the
Wigner function $W\left(\alpha\right)$ in Eq. (\ref{eq:wigner_monte_carlo})
respectively. Other parameters as in Fig (\ref{fig:fringes_plot_upperbound}).
Here, the optomechanical cat state decoheres after\textcolor{black}{{}
a storage time of $0.3466/\Gamma_{m}$, which is the time a Wigner
function loses its negativity according to Eq. (\ref{eq:upper_t_wigner_positive}).
\label{fig:wigner_upperbound}}}
\end{figure}

The Wigner function and its projection onto the phase space plane
are plotted in Fig. \ref{fig:wigner_upperbound-1} and Fig. \ref{fig:wigner_upperbound}
for storage times $0.02/\Gamma_{m},0.3466/\Gamma_{m}$ respectively.
The storage time $0.3466/\Gamma_{m}$ corresponds to the time a Wigner
function loses its negativity for a mean mechanical thermal number
$\bar{n}_{th}=0$ as given by Eq. (\ref{eq:upper_t_wigner_positive}).

\subsection{Reconstructed density operator }

\begin{figure}[H]

\includegraphics[width=0.9\columnwidth]{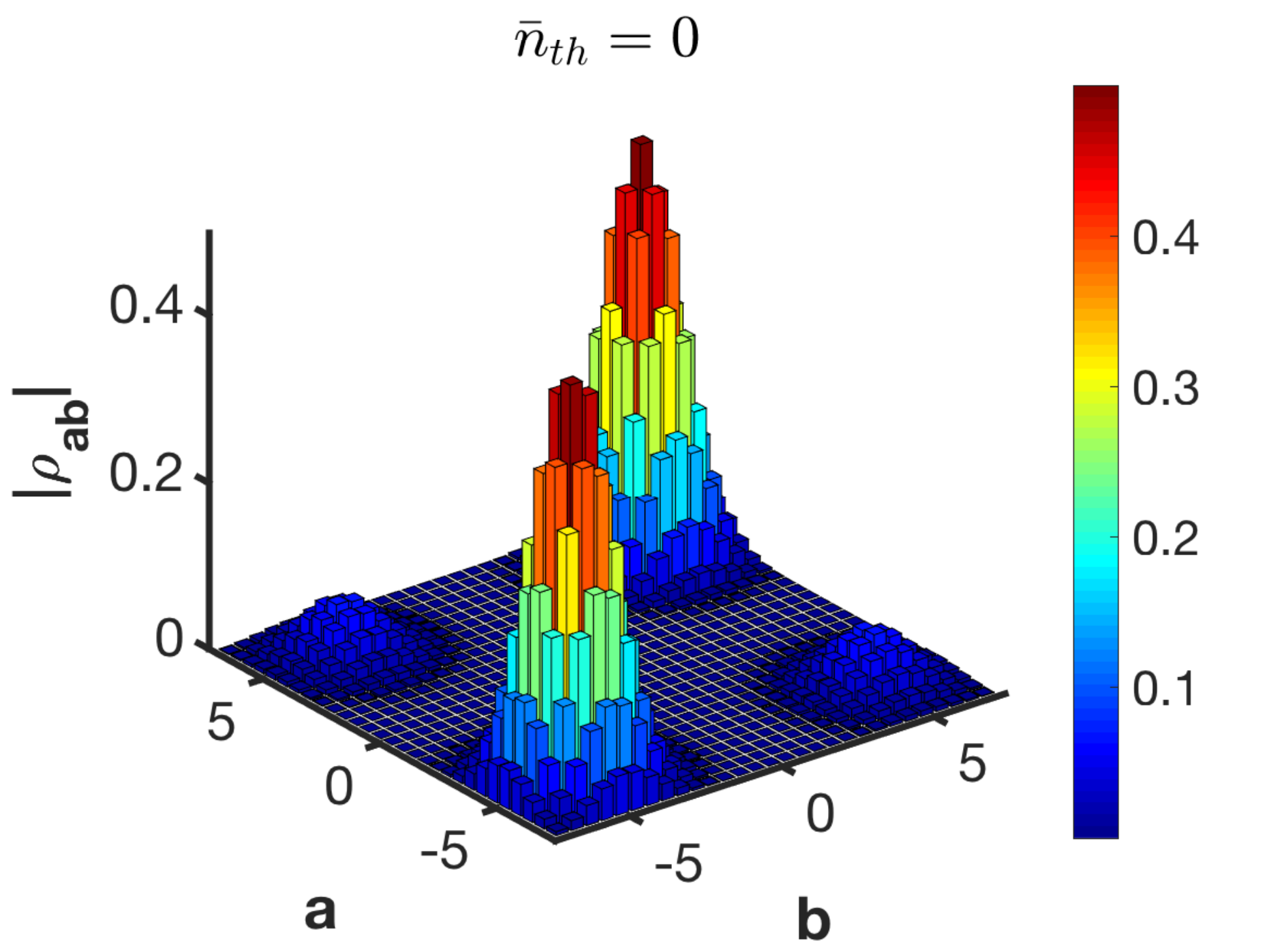}

\caption{The reconstructed density operator computed using the positive-P
distribution and Eq. (\ref{eq:density_op_reconstruction-1}) for $\alpha_{0}=5$
after reading out from the quantum memory. The mean mechanical thermal
noise and internal loss rate are chosen to be $\bar{n}_{th}=\Gamma_{int}=0$
, and the\textcolor{black}{{} storage time is $0.02/\Gamma_{m}$. }A
total number of $4$ samples are taken.\textcolor{black}{{} \label{fig:densityoperator_upperbound-1}}}
\end{figure}

\begin{figure}[H]

\includegraphics[width=0.9\columnwidth]{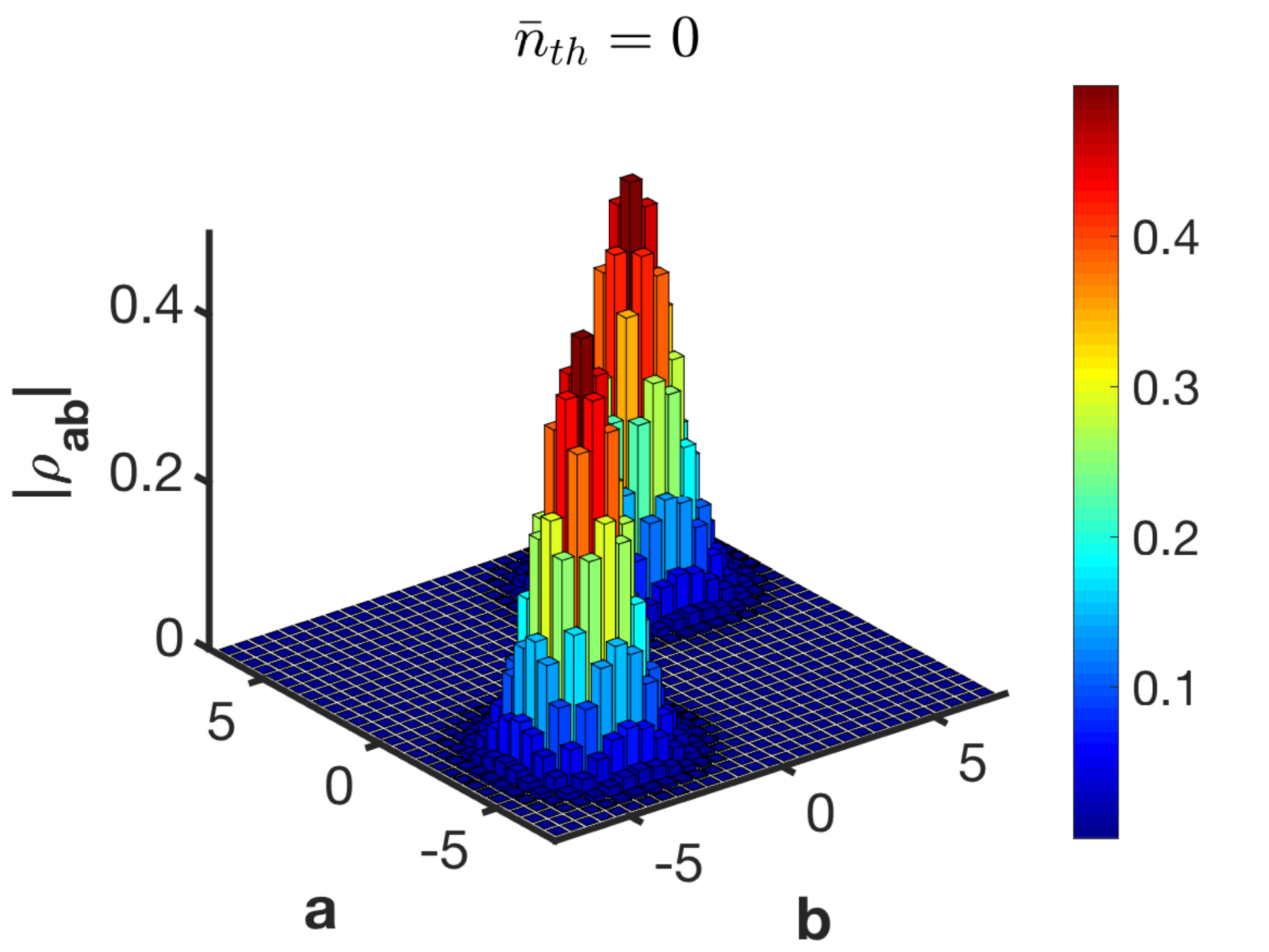}

\caption{The reconstructed density operator computed using the positive-P
distribution and Eq. (\ref{eq:density_op_reconstruction-1}) for $\alpha_{0}=5$
after reading out from the quantum memory. The mean mechanical thermal
noise and internal loss rate are chosen to be $\bar{n}_{th}=\Gamma_{int}=0$
. Here, the optomechanical cat state decoheres due to the finite mechanical
lifetime after\textcolor{black}{{} a storage time of $0.3466/\Gamma_{m}$,
which is the time a Wigner function loses its negativity according
to Eq. (\ref{eq:upper_t_wigner_positive}). }A total number of $4$
samples are taken.\textcolor{black}{{} \label{fig:densityoperator_upperbound}}}
\end{figure}

We reconstruct the density operator by looking at the modulus of the
density operator in the coherent state basis: $\left|\rho_{ab}\right|=\left|\langle a|\hat{\rho}_{0}^{out}|b\rangle\right|$.
This can be achieved using the Monte Carlo method as discussed in
the previous section and is shown below:

\begin{eqnarray}
\left|\rho_{ab}\right| & = & \left|\langle a|\hat{\rho}_{0}^{out}|b\rangle\right|\nonumber \\
 & \approx & \left|\frac{1}{N_{s}}\sum_{i}^{N_{s}}w\left(\vec{\alpha}_{0,i}^{in}\right)\frac{\langle a|\alpha_{0,i}^{out}\rangle\langle\alpha_{0,i}^{out+}{}^{*}|b\rangle}{\langle\alpha_{0,i}^{out+}{}^{*}|\alpha_{0,i}^{out}\rangle}\right|\,.\nonumber \\
\label{eq:density_op_reconstruction-1}
\end{eqnarray}
Here, the weight function $w$ is identical to the one given in Eq.
(\ref{eq:prob_dis_monte_carlo}). The reconstructed density operator
in the coherent state basis is plotted in Fig. \ref{fig:densityoperator_upperbound-1}
and Fig. \ref{fig:densityoperator_upperbound} for storage times $0.02/\Gamma_{m},0.3466/\Gamma_{m}$
respectively. The storage time $0.3466/\Gamma_{m}$ corresponds to
the time a Wigner function loses its negativity for a mean mechanical
thermal number $\bar{n}_{th}=0$ as given by Eq. (\ref{eq:upper_t_wigner_positive}).
Here we note the presence of the nonzero off-diagonal terms, for times
where the Wigner negativity is zero.

\subsection{Wigner negativity}

We also compute the Wigner negativity as defined in Eq. (\ref{eq:negative_volume_Wigner})
as a function of the optomechanical cat storage time and thermal noise.
The Wigner negativity can be easily computed numerically once the
Wigner function has been obtained, and we use the trapezoidal numerical
method to carry out the integration involved.

The numerical results are then compared with the corresponding analytical
results based on the idealized characteristic function solution in
Eq. (\ref{eq:p_ordered_chi_solution}). We define an auxiliary amplitude
given by
\begin{equation}
\alpha_{\pm}\left(t\right)=\alpha\pm\alpha_{0}e^{-\Gamma_{m}t}\,.\label{eq:aux_amplitude}
\end{equation}
The Wigner function at time $t$ as a function of cat state amplitude,
storage time and mean mechanical thermal number is given by 
\begin{eqnarray}
W\left(\alpha,t\right) & = & \frac{2}{\pi\mathcal{N}}\frac{1}{1+2\bar{n}_{th}\left(1-e^{-2\Gamma_{m}t}\right)}\times\nonumber \\
 &  & \left\{ exp\left[-\frac{2\alpha_{-}^{*}\left(t\right)\alpha_{-}\left(t\right)}{1+2\bar{n}_{th}\left(1-e^{-2\Gamma_{m}t}\right)}\right]\right.\nonumber \\
 &  & +exp\left[-\frac{2\alpha_{+}^{*}\left(t\right)\alpha_{+}\left(t\right)}{1+2\bar{n}_{th}\left(1-e^{-2\Gamma_{m}t}\right)}\right]\nonumber \\
 &  & +\langle\alpha_{0}|-\alpha_{0}\rangle exp\left[-\frac{2\alpha_{-}^{*}\left(t\right)\alpha_{+}\left(t\right)}{1+2\bar{n}_{th}\left(1-e^{-2\Gamma_{m}t}\right)}\right]\nonumber \\
 &  & \left.+\langle-\alpha_{0}|\alpha_{0}\rangle exp\left[-\frac{2\alpha_{+}^{*}\left(t\right)\alpha_{-}\left(t\right)}{1+2\bar{n}_{th}\left(1-e^{-2\Gamma_{m}t}\right)}\right]\right\} \,.\nonumber \\
\label{eq:Wigner_cat_state_time}
\end{eqnarray}
\begin{figure}[H]
\includegraphics[width=0.9\columnwidth]{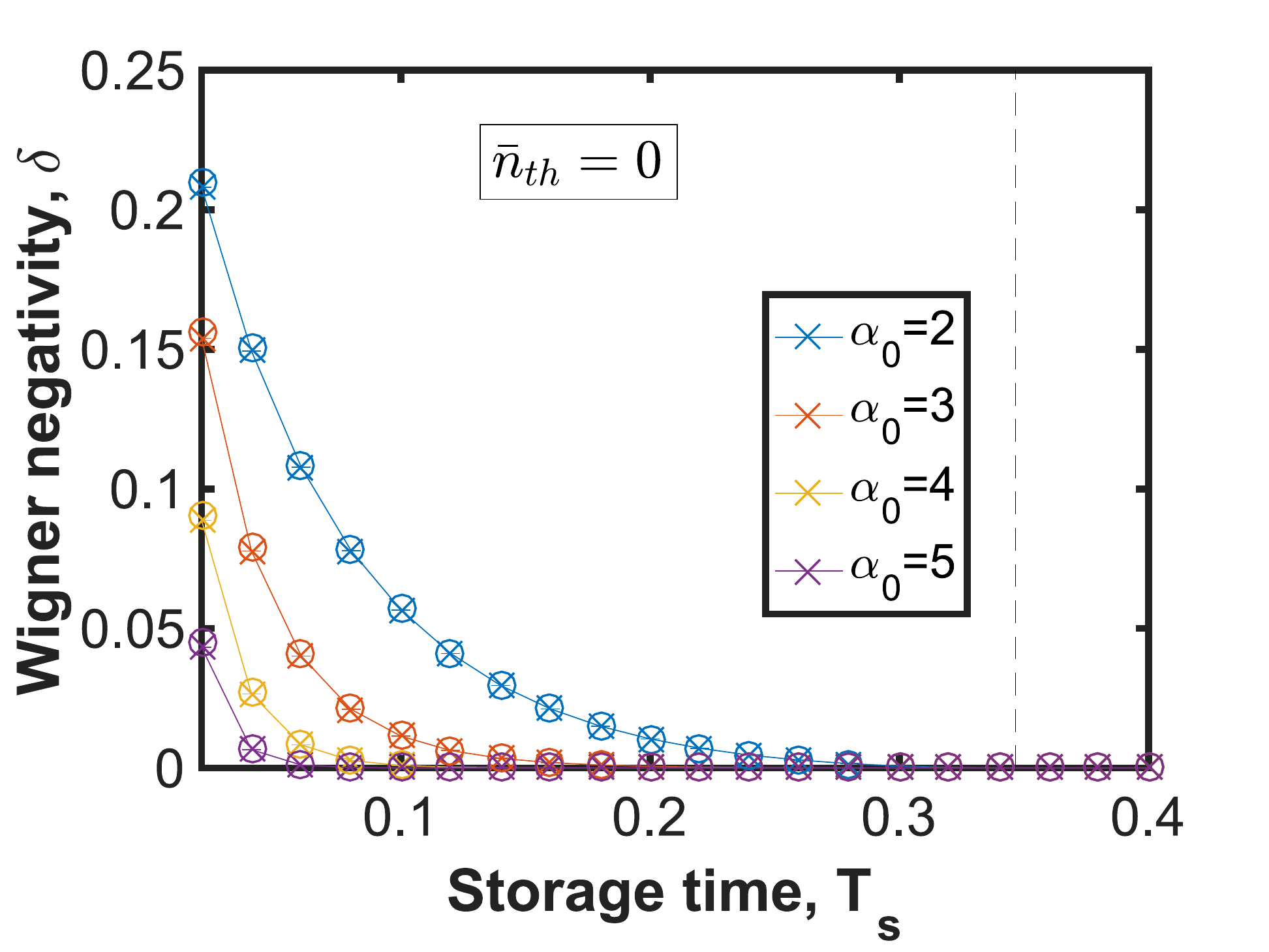}

\caption{The Wigner negativity of the read-out state as a function of the dimensionless
storage time (in multiples of $1/\Gamma_{m}$) for cat amplitudes
$\alpha_{0}=2,3,4$ and $5$. The mean mechanical thermal occupation
number and internal loss rate are chosen to be $\bar{n}_{th}=\Gamma_{int}=0$.
\textcolor{black}{The corresponding data points in circles are analytical
values based on Eq. (\ref{eq:Wigner_cat_state_time}). T}he dashed
vertical line is the upper bound of the time for a Wigner function
to lose its negativity, as given in Eq. (\ref{eq:upper_t_wigner_positive}).
For $\bar{n}_{th}=0$, the upper bound, in multiples of $1/\Gamma_{m}$,
is $0.3466$.\textcolor{red}{{} }A total number of $4$ samples are
taken. The error bars denote the time-step error in the phase-space
simulations. \textcolor{red}{\label{fig:wigner_negative_result}}}
\end{figure}
The Wigner negativity from both the analytical and numerical methods
are plotted in Fig. \ref{fig:wigner_negative_result} and Fig. \ref{fig:wigner_negative_result-1}
for a mean mechanical thermal occupation number $\bar{n}_{th}$ of
$0$ and $2$, respectively. Fig. \ref{fig:wigner_negativity_parameter_space}
shows a three-dimensional representation of the Wigner negativity
results as a function of mean mechanical thermal occupation number
and storage time.

\subsection{Variance of the $p$-quadrature}

Here, we compute the variance of $p$-quadrature before the cat-state
is stored and after the state has been read out from the quantum memory.
In particular, we compute this observable for storage times where
the corresponding Wigner functions for the quantum memory output states
lose their negativity, with zero mean mechanical thermal number. Note
that the positive-P representation computes normally ordered observables.
Hence, a quantity such as $\langle\hat{p}^{2}\rangle$ has to be normally
ordered first for the numerical results in the positive-P representation
to be correct. Thus
\begin{align}
\langle\hat{p}^{2}\rangle & =-\frac{1}{2}\left(\langle\hat{a}^{2}\rangle+\langle\hat{a}^{\dagger2}\rangle-2\langle\hat{a}^{\dagger}\hat{a}\rangle-1\right)\nonumber \\
 & =-\frac{1}{2}\left(\langle\alpha^{2}\rangle_{p}+\langle\alpha^{+2}\rangle_{p}-2\langle\alpha^{+}\alpha\rangle_{p}-1\right)\,,\label{eq:posp_p}
\end{align}
where $\alpha,\alpha^{+}$ are the complex field amplitudes in the
positive-P representation. We compare the numerical results for the
variance with the corresponding analytical ones as given by Eq. (\ref{eq:varcat}).
The comparison is shown in Table \ref{tab:variance}.
\begin{figure}[H]
\includegraphics[width=0.9\columnwidth]{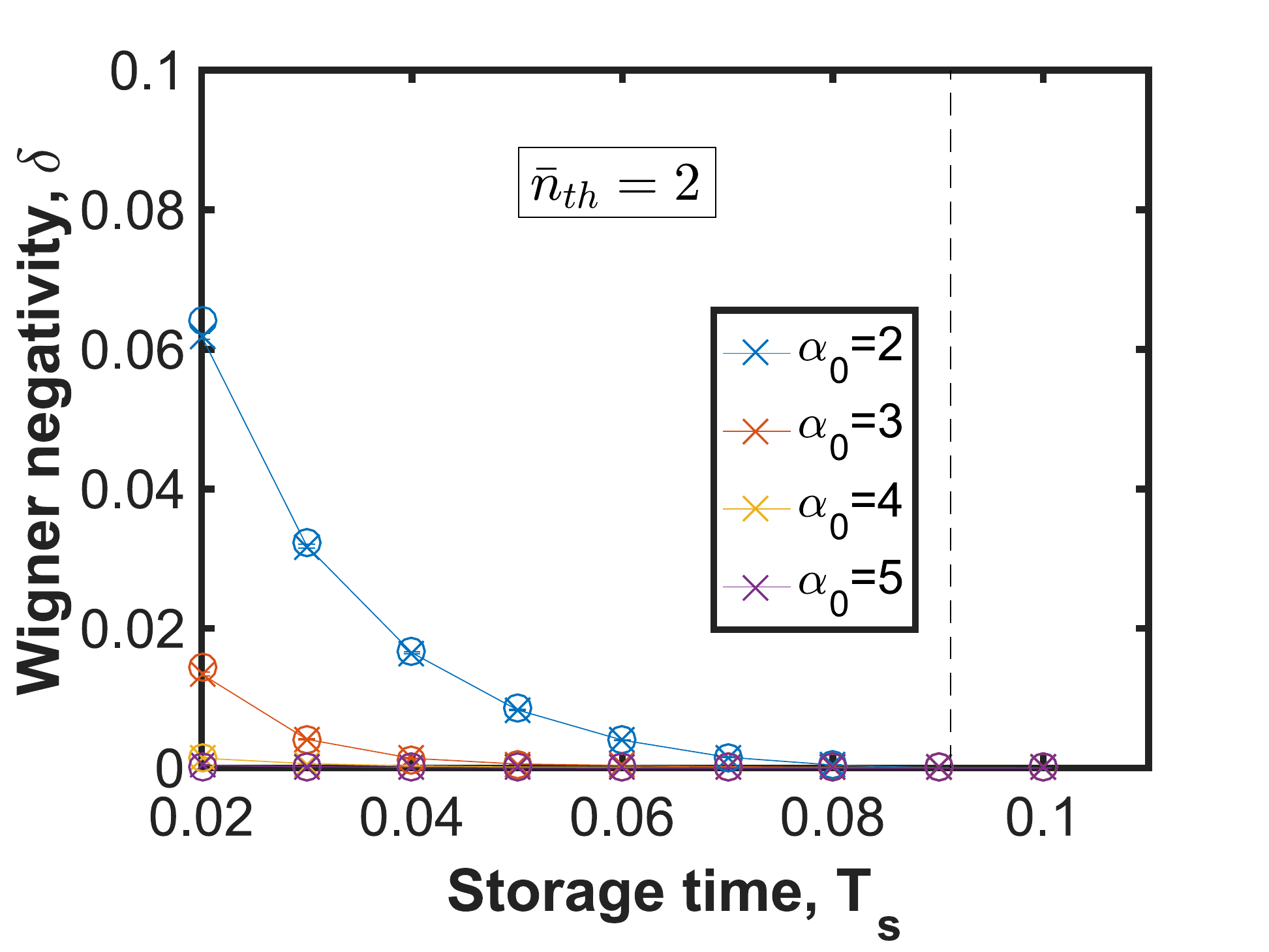}

\caption{The Wigner negativity of the read-out state as a function of the dimensionless
storage time (in multiples of $1/\Gamma_{m}$) for cat amplitudes
$\alpha_{0}=2,3,4$ and $5$. The mean mechanical thermal occupation
number $\bar{n}_{th}=2$, and the internal loss is $\Gamma_{int}=0$.
The dashed vertical line is the upper bound of the time for a Wigner
function to lose its negativity, as given in Eq. (\ref{eq:upper_t_wigner_positive}).
For $\bar{n}_{th}=2$, the upper bound, in multiples of $1/\Gamma_{m}$,
is $0.0912$. A total number of $2\times10^{5}$ samples are taken.\textcolor{red}{{}
}The error bars include both the sampling error and time-step error\textcolor{blue}{.}\textcolor{red}{\label{fig:wigner_negative_result-1}}
\textcolor{red}{}}
\end{figure}

\begin{figure}[H]
\includegraphics[width=0.9\columnwidth]{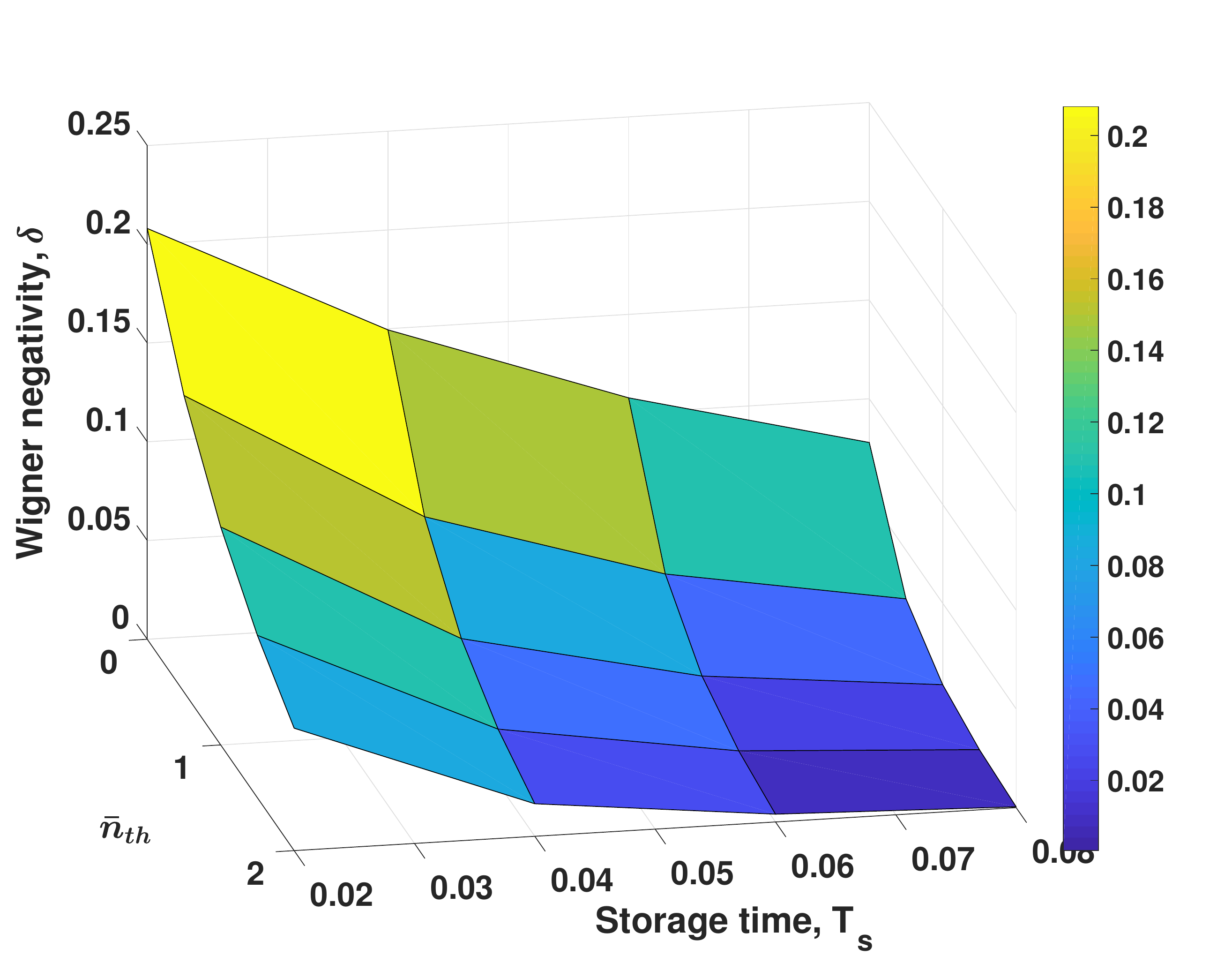}

\caption{The Wigner negativity of the read-out state as a function of the dimensionless
storage time, $T_{s}$ (in multiples of $1/\Gamma_{m}$) and the mean
mechanical thermal number $\bar{n}_{th}$ for a cat amplitude $\alpha_{0}=2$.
The internal loss is $\Gamma_{int}=0$. A total number of $2\times10^{5}$
samples are taken, except when $\bar{n}_{th}=0$, where $4$ samples
are taken instead. \textcolor{blue}{\label{fig:wigner_negativity_parameter_space}}}
\end{figure}

\begin{table}[H]
\begin{tabular}{|>{\centering}p{0.2\columnwidth}|>{\centering}p{0.25\columnwidth}|>{\centering}p{0.25\columnwidth}|>{\centering}p{0.25\columnwidth}|}
\hline 
Cat amplitude, $\alpha_{0}$ & Analytical prediction for a cat state, $\left(\Delta p\right)_{cat}^{2}$ & Numerical value before storage, $\left(\Delta p\right)_{in}^{2}$ & Numerical value after readout, $\left(\Delta p\right)_{out}^{2}$\tabularnewline
\hline 
\hline 
1 & 0.2616 & 0.2616 & 0.3809\tabularnewline
\hline 
2 & 0.4973 & 0.4973 & 0.4987\tabularnewline
\hline 
3 & 0.5000 & 0.5000 & 0.5000\tabularnewline
\hline 
5 & 0.5000 & 0.5000 & 0.5000\tabularnewline
\hline 
\end{tabular}

\caption{The analytical and numerical values for the variance of $p$-quadrature
for different cat amplitudes $\alpha_{0}$. The analytical values
are obtained using the expression in Eq. (\ref{eq:varcat}). These
values are obtained for the parameters $\bar{n}_{th}=0$ and a storage
time of $1/2\text{ln}\left(2\right)$, which is the upper bound time
for the loss of Wigner negativity of the readout state. \label{tab:variance}}
\end{table}

In practice, the variance of $p$-quadrature for a cat state is too
tiny to be differentiated from the variance of $p$-quadrature for
a mixed state, for a cat state amplitude larger than 2. However, in
the cases where $\left(\Delta p\right)^{2}<1/2$ can be observed,
the variance method serves as a sufficient criterion to verify the
existence of a cat state. This is crucial as we see that for $\bar{n}_{th}=0$
and a storage time that corresponds to a state where its Wigner function
loses its negativity, only the reconstructed density operator and
the variance methods are able to detect the presence of a density
operator with non-vanishing off-diagonal terms. The variance method
has the advantage that no state tomography is needed, as opposed to
the density operator reconstructed approach.

\subsection{Decoherence effects on an optomechanical cat state}

In the previous subsection, the internal cavity decay rate is set
to zero, which corresponds to an optimal optomechanical quantum state
transfer. In practice, the internal cavity decay rate is nonzero,
causing the quantum state transfer to be less efficient . This introduces
further decoherence to the quantum state that is stored. In this section,
we analyze more realistic parameter values that correspond to recent
electromechanical ecxperiments. 

\begin{figure}[H]
\includegraphics[width=0.9\columnwidth]{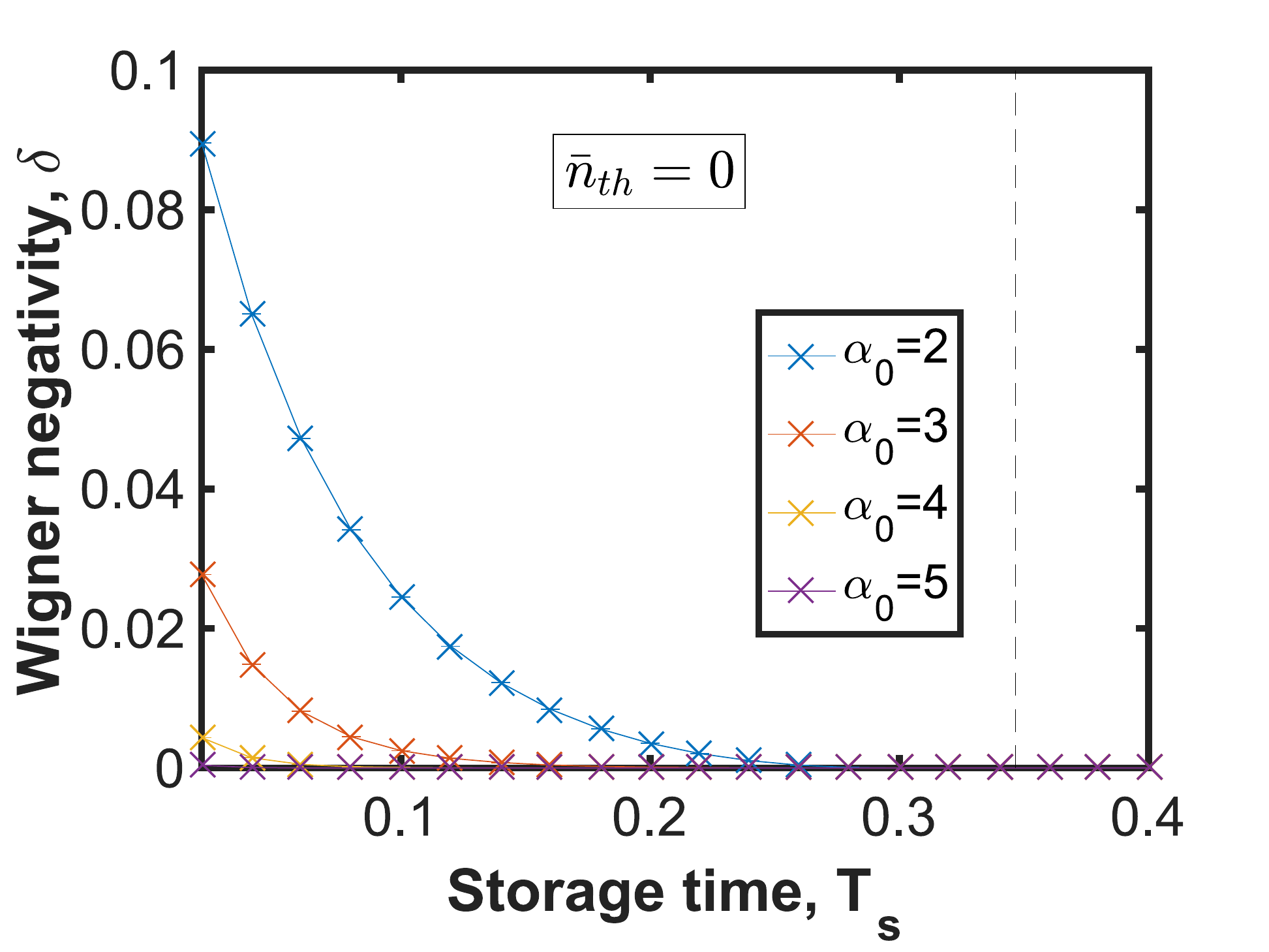}

\caption{The Wigner negativity of the read-out state as a function of the dimensionless
storage time (in multiples of $1/\Gamma_{m}$) for cat amplitudes
$\alpha_{0}=2,3,4$ and $5$. The mean mechanical thermal occupation
number $\bar{n}_{th}=0$. The internal cavity decay rate is nonzero
and contributes to further decoherence of the cat state. Here, the
internal cavity decay rate is set to be $\Gamma_{int}=0.05$. A total
number of $4$ samples are taken. The error bars denote the time-step
error in the phase-space simulations.\label{fig:extra_decoherence}}
\end{figure}

First we consider the case where there is a nonzero optical internal
loss, $\Gamma_{int}$. The state transfer protocol used in this paper
predicts that the stored amplitude, given an initial coherent amplitude
$\alpha$, would have an expectation value of 
\begin{align}
\langle b\left(0\right)\rangle & =\frac{\sqrt{2\Gamma_{ext}}G\alpha}{2\sqrt{\left(K_{+}+M\right)\left(K_{+}-M\right)K_{+}}}\,,\label{eq:mean_stored_amp}
\end{align}
based on Eq. (\ref{eq:stored_mode}). If we consider a realistic internal
cavity decay rate $\Gamma_{int}=0.05$, then from the set of parameters
we use, the stored amplitude is $0.9745\alpha$. As shown in Fig.
\ref{fig:extra_decoherence}, this significantly reduces the Wigner
negativity of the retrieved cat state, even at zero temperature. 

\begin{figure}[H]
\includegraphics[width=0.9\columnwidth]{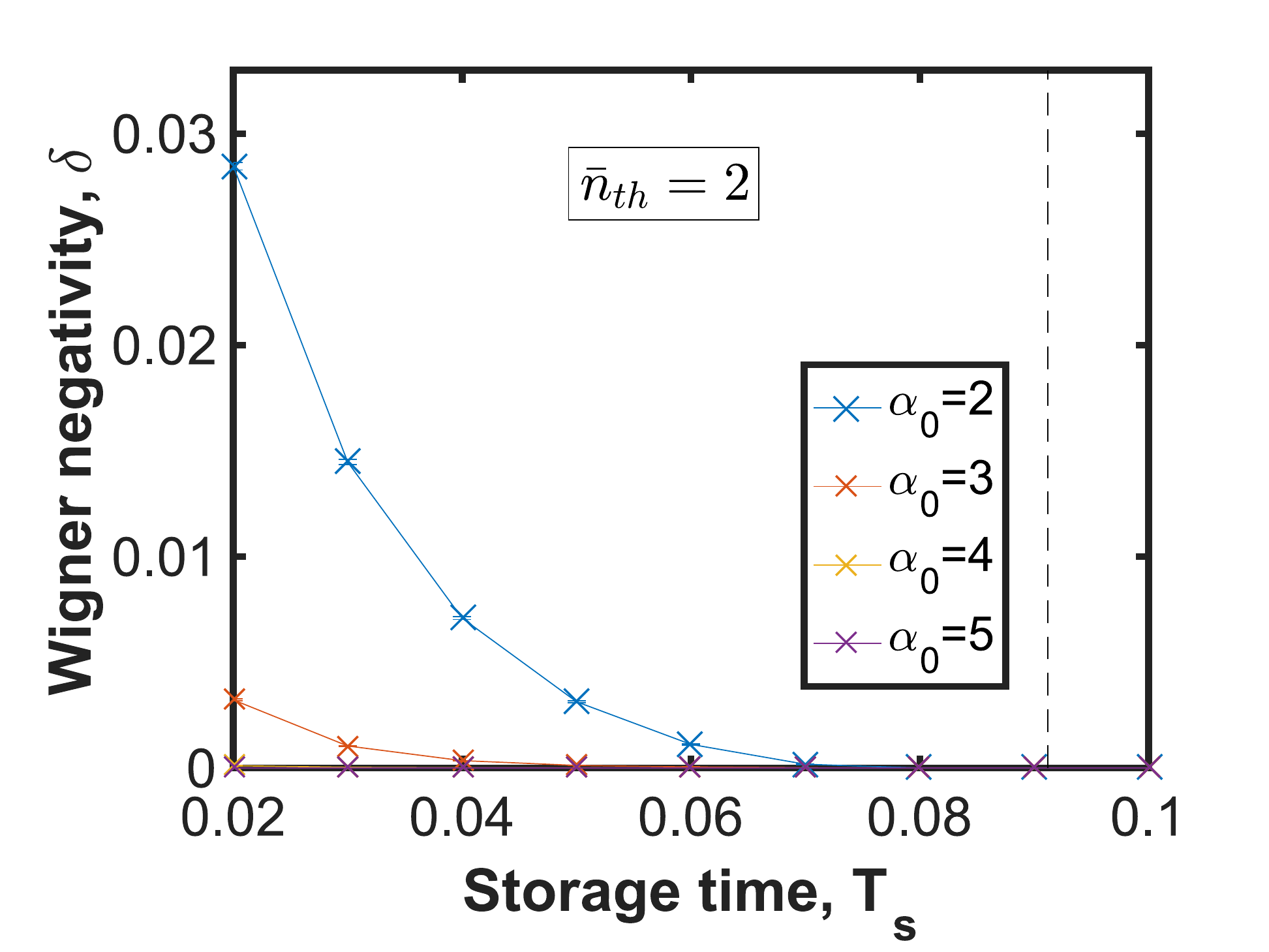}

\caption{The Wigner negativity of the read-out state as a function of the dimensionless
storage time (in multiples of $1/\Gamma_{m}$) for cat amplitudes
$\alpha_{0}=2,3,4$ and $5$. The mean mechanical thermal occupation
number $\bar{n}_{th}=2$. The internal cavity decay rate is nonzero
and contributes to further decoherence of the cat state. Here, the
internal cavity decay rate is set to be $\Gamma_{int}=0.05$ and the
initial mechanical mode has an occupation number of $0.5$. A total
number of $2\times10^{5}$ samples are taken. The error bars include
both the sampling error and time-step error. \label{fig:extra_decoherence-1} }
\end{figure}

Next, we consider the effect of finite thermal occupatoin numbers
in the mechanical mode. In Fig. \ref{fig:extra_decoherence-1}, we
show the result for the Wigner negativity for internal cavity decay
rate $\Gamma_{int}=0.05$ and mechanical thermal occupation number
$\bar{n}_{th}=2$. Also, the initial mechanical mode has an occupation
number of $0.5$, instead of being in its ground state, to give an
example of a possible non-ground-state initial condition.

With these more realistic parameter values, the maximum detectable
cat state has $\alpha_{0}=3$, with a squared separation of $S=\left|2\alpha_{0}^{2}\right|=36$.
This demonstrates that to store a mechanical cat state having $S=100$,
as has been generated experimentally in a microwave mode, will require
reductions in the loss rates and mechanical reservoir temperatures
compared to currently achieved values.

The ratio between the external cavity decay rate $\Gamma_{ext}$ and
the total cavity decay rate $\Gamma_{o}$ has been quoted as the efficiency
of an optomechanical state transfer protocol \citep{palomaki2013coherent}.
In our case, $\Gamma_{ext}/\Gamma_{o}=0.95$, and we note that this
only quantifies the amplitude being stored; the coherent quantum superposition
in the quantum state has to be stored too. A quantum memory that has
high amplitude efficiency, while retaining the quantum superposition
of the stored quantum state is a challenge. The detection inefficiency
which is not included in our model will no doubt make the verification
of nonclassical quantum states even more difficult \citep{Skotiniotis2017macroscopic}.
However, with the improvement in technologies such as optomechanical
cooling using squeezed states \citep{clark2017sideband}, efficient
quantum state transfer \citep{Wang1087,Reed:2017aa} and detection
schemes, the generation and verification of optomechanical cat states
becomes feasible. 

\section{conclusion}

In summary, we analyze a protocol for optomechanical storage of a
Schr\"odinger cat state. To analyze its properties, a simplified
decoherence model for a stored cat state was investigated by solving
the single-mode master equation analytically. Additionally, the full
coupled system including input and output was simulated using the
positive-P phase space method. Provided importance sampling is utilized,
this provides a compact and efficient probabilistic representation
of such macroscopic quantum superpositions. The method allows straightforward
quantum state sampling to be carried out, even for these highly nonclassical,
entangled multimode transients. 

We then discussed typical cat state signatures as a measure of the
quality of the quantum memory, and described the numerical methods
required to compute these cat state signatures. The analytical predictions
of the simplified model were then compared with our numerical results,
showing good agreement. With the advent of finer quantum controls
and manipulations in optomechanics and their physical implementations
in different systems, the goal of creating and storing a small optomechanical
cat state does appear achievable. We have investigated a number of
different sources of decoherence, including losses in the optical
system, losses in the mechanical system, initial thermal occupation
of the mechanical oscillator, and finite temperature mechanical reservoirs.
All of these clearly play a role in reducing the cat-state signatures,
especially as the stored photon number is increased, but are not an
insuperable barrier.

Our numerical methods provide an efficient way to probe the feasibility
of this protocol with realistic experimental parameters. We show that
a moderate size Schr\"odinger cat state with $n\le9$ stored quanta
and a phase-space squared separation of $S=36$ appears feasible with
present quantum technologies.

\section*{Acknowledgements}

MDR acknowledges support from Australian Research Council Discovery
Grant DP180102470. PDD and MDR thank the hospitality of the Institute
for Atomic and Molecular Physics (ITAMP) at Harvard University, supported
by the NSF. 

\section*{Appendix}

\subsection*{Decoherence of the cat state}

A cat state is extremely sensitive to fluctuations and losses due
to the interaction with its environment. Here, we assume a simple
model of decoherence provided by a master equation that includes damping
and thermal noise, in order to obtain an analytical solution for the
time evolution of a cat state in a simple \emph{gedanken-experiment}.
The time it takes for the Wigner function of a cat state to become
positive is also investigated.  This gives analytical insight and
provides a comparison for the numerical results of the main text,
which compute the final readout cat-state after a storage time in
a quantum memory.

The time evolution of a single-mode density operator due to its interaction
with a lossy environment is given by the following master equation:
\begin{eqnarray}
\frac{\partial}{\partial t}\hat{\rho} & = & \gamma\bar{n}\left(2a^{\dagger}\hat{\rho}a-aa^{\dagger}\hat{\rho}-\hat{\rho}aa^{\dagger}\right)\nonumber \\
 &  & +\gamma\left(\bar{n}+1\right)\left(2a\hat{\rho}a^{\dagger}-a^{\dagger}a\hat{\rho}-\hat{\rho}a^{\dagger}a\right)\,.\label{eq:master_eqn}
\end{eqnarray}
Here, $\hat{\rho}$ is the cat state density operator, $\gamma$ is
the decay rate of the relevant mode and $\bar{n}$ is the average
thermal occupation number due to the interaction with the environment.
Using phase space methods, we transform the above master equation
into a time evolution equation of an $s$-ordered characteristic function.
The advantage of using phase space methods is that the corresponding
equations are much easier to solve than the operator equation Eq.
(\ref{eq:master_eqn}). Here, the $s$-ordered characteristic function
is based on the definition by Cahill and Glauber \citep{PhysRev.177.1882}
and is given by
\begin{align}
\chi_{s}\left(\lambda\right) & =\text{Tr}\left[\hat{\rho}e^{\lambda\hat{a}^{\dagger}-\lambda^{*}\hat{a}+s\left|\lambda\right|^{2}/2}\right]\,,\label{eq:s-ordered_characteristic_function}
\end{align}
such that $s=-1,0,1$ corresponds to the characteristic function in
Q, Wigner and P representations, respectively. By multiplying both
sides of the Eq. (\ref{eq:master_eqn}) by $\text{e}^{\lambda\hat{a}^{\dagger}}\text{e}^{-\lambda^{*}\hat{a}}$
and taking the trace, it can be shown that the $s$-ordered characteristic
function satisfies the following time evolution equation \citep{barnett1997methods}:
\begin{eqnarray}
\frac{\partial}{\partial t}\chi_{s}\left(\lambda,t\right) & = & -\gamma\left(\lambda\frac{\partial}{\partial\lambda}+\lambda^{*}\frac{\partial}{\partial\lambda^{*}}\right)\chi_{s}\nonumber \\
 &  & -\gamma\left[s-\left(2\bar{n}+1\right)\right]\left|\lambda\right|^{2}\chi_{s}\,.\label{eq:characteristic_equation_time_evolution}
\end{eqnarray}
Eq. (\ref{eq:characteristic_equation_time_evolution}) can be solved
analytically using the method of characteristics. These analytical
solutions allow us to compare with the numerical solutions obtained
from a full quantum simulation in later sections. 

Since characteristic functions of different order are related, we
may choose $\bar{s}=2\bar{n}+1$ to simplify the partial differential
equation Eq. (\ref{eq:characteristic_equation_time_evolution}). The
corresponding partial differential equation is
\begin{eqnarray}
\frac{\partial}{\partial t}\chi_{\bar{s}}\left(\lambda,t\right) & = & -\gamma\left(\lambda\frac{\partial}{\partial\lambda}+\lambda^{*}\frac{\partial}{\partial\lambda^{*}}\right)\chi_{\bar{s}}\nonumber \\
\label{specific_order_chi_time_evolution}
\end{eqnarray}
and the solution can be shown to be \citep{barnett1997methods}
\begin{eqnarray}
\chi_{\bar{s}}\left(\lambda,t\right) & = & \chi_{\bar{s}}\left(\lambda e^{-\gamma t},0\right)\,.\nonumber \\
\label{eq:characteristic_function_solution}
\end{eqnarray}
The $s$-ordered characteristic function at time $t$ is then obtained
through the relation 
\begin{eqnarray}
\chi_{s}\left(\lambda\right) & = & \text{exp}\left\{ -\left[\bar{s}-s\right]\frac{\left|\lambda\right|^{2}}{2}\right\} \chi_{\bar{s}}\,.\nonumber \\
\label{eq:relation_different_ordered_chi}
\end{eqnarray}
Using Eq. (\ref{eq:relation_different_ordered_chi}) and the solution
of the characteristic function in Eq. (\ref{eq:characteristic_function_solution}),
the solution of an $s$-ordered characteristic function at time $t$
is given by
\begin{eqnarray}
\chi_{s}\left(\lambda,t\right) & = & \text{exp}\left\{ -\left[\bar{s}-s\right]\frac{\left|\lambda\right|^{2}}{2}\left(1-e^{-2\gamma t}\right)\right\} \nonumber \\
 &  & \times\chi_{s}\left(\lambda e^{-\gamma t},0\right)\,.\label{eq:p_ordered_chi_solution}
\end{eqnarray}

Based on this solution of the master equation, we can now investigate
the time it takes for a cat state to lose its coherence. In the formalism
of density operators, this corresponds to the absence of off-diagonal
elements in a density operator. The corresponding density operator
then describes a statistical mixture of two coherent states. 

In the following subsections, we first compute the time taken for
the off-diagonal terms of a cat density operator to vanish, when expressed
using a coherent state basis. Another way to characterize the nonclassicality
of a cat state is the negativity of the Wigner function. We also derive
the upper bound for the time it takes for the Wigner function of a
cat state to become positive. 

\subsection*{Density operator off-diagonal terms: zero temperature case }

In this subsection, we consider the case where the environment is
at zero temperature $T=0$, so that the mean mechanical thermal occupation
number $\bar{n}_{th}=0$. In this limit, the decay of the cat-state
quantum coherence is due to the finite quantum memory decay rate.
This allows us to gain insight on the rate of cat-state decoherence.
The normally ordered characteristic function for the cat density operator
(\ref{eq:cat_state_density_op}), $\chi_{1}\left(\lambda\right)$,
is a sum of four terms:
\begin{align}
\chi_{1}\left(\lambda\right) & =\frac{1}{\mathcal{N}}\left[e^{\lambda\alpha_{0}^{*}}e^{-\lambda^{*}\alpha_{0}}+e^{-\lambda\alpha_{0}^{*}}e^{\lambda^{*}\alpha_{0}}\right.\label{eq:initial_cat_char_function}\\
 & \left.+\langle-\alpha_{0}|\alpha_{0}\rangle e^{-\lambda\alpha_{0}^{*}}e^{-\lambda^{*}\alpha_{0}}+\langle\alpha_{0}|-\alpha_{0}\rangle e^{\lambda\alpha_{0}^{*}}e^{\lambda^{*}\alpha_{0}}\right]\,.\nonumber 
\end{align}
Here, the first two terms correspond to the diagonal elements of the
cat density operator and the last two terms correspond to the off-diagonal
terms.

Next, we obtain the expression for the characteristic function of
a cat state at time $t$, $\chi_{s}\left(\lambda,t\right)$. From
Eq. (\ref{eq:p_ordered_chi_solution}) and further setting $s=1$
(which corresponds to the normally ordered characteristic function),
we find an expression with four terms involving exponentials of $\alpha_{0}$,
$\alpha_{0}^{*}$ and $\left|\alpha_{0}\right|^{2}$, together with
time-dependent factors. We identify two terms as the diagonal terms
in a density operator $|\alpha_{0}e^{-\gamma t}\rangle\langle\alpha_{0}e^{-\gamma t}|$
and $|-\alpha_{0}e^{-\gamma t}\rangle\langle-\alpha_{0}e^{-\gamma t}|$
respectively, and the other two terms correspond to the off-diagonal
terms $|\alpha_{0}e^{-\gamma t}\rangle\langle-\alpha_{0}e^{-\gamma t}|$
and $|-\alpha_{0}e^{-\gamma t}\rangle\langle\alpha_{0}e^{-\gamma t}|$
respectively, with a time dependent coefficient $e^{-2|\alpha_{0}|^{2}(1-e^{-2\gamma t})}$.
The resulting density operator is given by 
\begin{align}
\hat{\rho}_{cat}\left(t\right) & =\frac{1}{\mathcal{N}}\left[|\alpha_{0}e^{-\gamma t}\rangle\langle\alpha_{0}e^{-\gamma t}|+|-\alpha_{0}e^{-\gamma t}\rangle\langle-\alpha_{0}e^{-\gamma t}|\right.\nonumber \\
 & +e^{-2|\alpha_{0}|^{2}(1-e^{-2\gamma t})}|\alpha_{0}e^{-\gamma t}\rangle\langle-\alpha_{0}e^{-\gamma t}|\nonumber \\
 & \left.+e^{-2|\alpha_{0}|^{2}(1-e^{-2\gamma t})}|-\alpha_{0}e^{-\gamma t}\rangle\langle\alpha_{0}e^{-\gamma t}|\right]\,.\label{eq:cat_t}
\end{align}
The off-diagonal terms in Eq. (\ref{eq:cat_t}) vanish in a shorter
time for larger coherent amplitude $\alpha_{0}$ and damping rate
$\gamma$. We note that in the absence of thermal noise, the off-diagonal
terms never completely vanish i.e. there is no ``sudden death''
effect of the type discussed in Ref. \citep{PhysRevA.84.012121}.

\subsection*{Negativity of the Wigner function: finite temperature case }

Here, we derive the upper bound on the time it takes for the cat state
Wigner function to become completely positive. In this subsection,
we include the effect of thermal noise. This approach is based on
the paper of Paavola et al. \citep{PhysRevA.84.012121}. In that paper,
the upper bound of the time for any P function to lose its negativity
is obtained by calculating the condition for that initial P function
to turn into a Q function, which is always positive. The upper bound
$t_{p}$ was found to be 
\begin{align}
t_{p} & =\frac{1}{2\gamma}\text{ln}\left(\frac{1}{\bar{n}_{th}}+1\right)\,,\label{eq:p_function_upper_bound}
\end{align}
where $\bar{n}_{th}$ is the mean mechanical thermal occupation number
and $\gamma$ is the decay rate of the system. Following the same
method, we obtain the upper bound of the time for a cat Wigner function
to lose its negativity.

The Wigner function at time $t$ is given by: 
\begin{align}
W\left(\alpha,t\right) & =\intop\chi_{-1}\left(\lambda,t\right)e^{|\lambda|^{2}/2}e^{\lambda^{*}\alpha}e^{-\lambda\alpha}\,\frac{d^{2}\lambda}{\pi^{2}}\\
 & =\intop\chi_{-1}\left(\lambda e^{-\gamma t},0\right)e^{q(t)\left|\lambda\right|^{2}+\alpha(\lambda^{*}-\lambda)}\,\frac{d^{2}\lambda}{\pi^{2}},\nonumber 
\end{align}
 where Eq. (\ref{eq:p_ordered_chi_solution}) is used in the second
line, and $q(t)\equiv1/2-\left(1+\bar{n}_{th}\right)\left(1-e^{-2\gamma t}\right)$. 

The right side of the equation above will correspond to a Q function,
which is always positive, if the  condition $q(t)=0$. The upper
bound for the time it takes for the Wigner function of the cat state
to be positive $t_{+}$ is therefore
\begin{align}
t_{+} & =\frac{1}{2\gamma}\text{ln}\left(\frac{1+\bar{n}_{th}}{\frac{1}{2}+\bar{n}_{th}}\right)\,.\label{eq:upper_t_wigner_positive}
\end{align}
Note that $t_{+}$ is not the time where a cat Wigner function is
always positive, but the \emph{upper bound} for the time it takes
for a cat Wigner function to become positive. It is a function of
the damping rate and the expectation value of the thermal occupation
number, and is not a function of the size of the cat state.

To this end, it is worth noting that a non-negative Wigner function
does \emph{not} imply there is no cat-state quantum coherence. The
numerical results for other cat-state signatures calculated at the
time corresponding to $t_{+}$ are given in Section \ref{sec:Numerical-simulation-and}.
At the time $t_{+}$, while the Wigner negativity is zero, other signatures
can indicate the presence of a cat-state. Let us focus on the density
operator in Eq. (\ref{eq:cat_t}) at the time $t_{+}$ for $\bar{n}_{th}=0$.
At the time $t_{+}=1/2\gamma\text{ln}2$, the off-diagonal terms in
the density operator Eq. (\ref{eq:cat_t}) do \emph{not} vanish, albeit
they make a tiny contribution that scales with the cat state amplitude
as $exp\left(-\left|\alpha_{0}\right|^{2}\right)$. This suggests
that more than one signature should be measured and calculated in
an experiment to conclusively verify the existence of a cat state.

\appendix
\bibliographystyle{apsrev4-1}
\bibliography{cat_transfer_references}

\end{document}